\documentclass{article}
\usepackage{arxiv}

\usepackage{authblk}
\usepackage{hyperref}
\usepackage[utf8]{inputenc}
\usepackage[T1]{fontenc}
\usepackage{graphicx}
\usepackage{booktabs}
\usepackage{multirow}

\usepackage{subfig}

\usepackage[linesnumbered, vlined, ruled]{algorithm2e}
\usepackage{xcolor}
\usepackage{mathtools}
\usepackage{amsfonts}
\usepackage{amsmath}
\usepackage{fixmath}

\newtheorem{mydefinition}{Definition}

\newtheorem{myexample}{Example}

\newcommand{\sbtsr}{\mbox{$\mathcal{S}$BTSR-tree}\xspace}
\newcommand{\btsr}{\mbox{BTSR-tree}\xspace}

\usepackage{soul,xcolor}
\date{}

\begin{document}

\title{Local Similarity Search on Geolocated Time Series Using Hybrid Indexing}

\author[1]{Georgios Chatzigeorgakidis}
\author[2]{Dimitrios Skoutas}
\author[3]{Kostas Patroumpas}
\author[4]{Themis Palpanas}
\author[5]{Spiros Athanasiou}
\author[6]{Spiros Skiadopoulos}

\affil[1]{IMSI, Athena R.C., Greece, \textit{email: gchatzi@athenarc.gr}}
\affil[2]{IMSI, Athena R.C., Greece, \textit{email: dskoutas@athenarc.gr}}
\affil[3]{IMSI, Athena R.C., Greece, \textit{email: kpatro@athenarc.gr}}
\affil[4]{LIPADE, Paris Descartes University, France, French University Institute (IUF) \textit{email: themis@mi.parisdescartes.fr}}
\affil[5]{IMSI, Athena R.C., Greece, \textit{email: spathan@athenarc.gr}}
\affil[6]{Dept. of Inf. \& Telecommunications University of Peloponnese, Greece \textit{email: spiros@uop.gr}}

\maketitle

\begin{abstract}
    Geolocated time series, i.e., time series associated with certain locations, abound in many modern applications. In this paper, we consider hybrid queries for retrieving geolocated time series based on filters that combine spatial distance and time series similarity. For the latter, unlike existing work, we allow filtering based on local similarity, which is computed based on subsequences rather than the entire length of each series, thus allowing the discovery of more fine-grained trends and patterns. To efficiently support such queries, we first leverage the state-of-the-art \btsr index, which utilizes bounds over both the locations and the shapes of time series to prune the search space. Moreover, we propose optimizations that check at specific timestamps to identify candidate time series that may exceed the required local similarity threshold. To further increase pruning power, we introduce the \sbtsr index, an extension to \btsr, which additionally segments the time series temporally, allowing the construction of tighter bounds. Our experimental results on several real-world datasets demonstrate that \sbtsr can provide answers much faster for all examined query types. This paper has been published in the 27\textsuperscript{th} International Conference on Advances in Geographic Information Systems (ACM SIGSPATIAL 2019).
\end{abstract}

\section{Introduction}
\label{sec:intro}

A {\em time series} is a time-ordered sequence of data points. Time series are ubiquitous in many application domains. They can represent various types of measurements, such as user check-ins at various Points of Interest, energy consumption in smart buildings, PM2.5 particle concentration measured by air pollution sensors, etc. Analyzing and mining time series data is highly important for discovering trends and patterns in such phenomena, and has attracted extensive research interest over the last years \cite{DBLP:journals/pvldb/EchihabiZPB18,DBLP:conf/sigmod/LinardiZPK18a,DBLP:journals/datamine/YehZUBDDZSMK18}.

However, what is usually overlooked is that the phenomena represented by time series are often also associated with geographic locations, e.g., time series generated by sensors installed at fixed positions. In such cases, spatial distance also plays an important role in the analysis, since discovery of trends and patterns may depend not only on time series similarity but also on geographic proximity. Motivated by this observation, in previous work \cite{chatzig17btsr, DBLP:conf/gis/Chatzigeorgakidis18} we introduced the concept of {\em geolocated} time series and we  proposed hybrid indexing techniques that efficiently support the retrieval of time series based on both spatial distance and time series similarity.

In particular, we introduced the \btsr \cite{chatzig17btsr}, a {\em hybrid index} that first builds an R-tree over the locations of the time series data. It then enhances each node with appropriate upper- and lower-bounding time series (MBTS) that enclose the subset of time series represented by it. Combining MBTSs and MBRs, the query evaluation algorithm can simultaneously prune the search space based on time series similarity and spatial distance while traversing the index. To further increase its pruning power, the \btsr groups together similar time series within each node to derive tighter bounds.

This existing approach for hybrid search over geolocated time series using the \btsr supports only {\em global} time series similarity, i.e., similarity measured across the entire length of time series. Specifically, as in other works in this area \cite{DBLP:journals/pvldb/EchihabiZPB18,jessica2007dmkd,camerra2010icdm,camerra2014kais}, the distance between two time series is measured by aggregating the pairwise Euclidean distance of their respective values across the entire sequences. However, in many cases, more fine-grained trends and patterns may exist, which are missed under this global similarity measure. For example, consider two time series representing the hourly energy consumption of two nearby buildings over a week, and assume that the two buildings exhibit a similar consumption pattern during working days but a different one in weekends. A query imposing a similarity threshold over the entire week would fail to identify these two geolocated time series as similar. However, it may be useful to discover that there is a period of up to 5 days during which these two time series are actually similar.



Motivated by this observation, in this work we extend our previous approach on hybrid queries over geolocated time series to support {\em local similarity} of time series, thus allowing more flexible and fine-grained queries and analyses.
The {\em local similarity score} between two time series $T_i$ and $T_j$ is defined as the maximum number of consecutive timestamps during which the respective values of $T_i$ and $T_j$ do not differ by more than a user-specified threshold $\epsilon$.
Notice that, compared to global similarity, this condition is more relaxed, in the sense that it is applied to subsequences of length lower than $T_i$ and $T_j$, but at the same time stricter, in the sense that the threshold $\epsilon$ is required to be satisfied at each individual timestamp during the selected period rather than on the aggregate distance over all timestamps.

\begin{figure}[!t]
 \centering
 \includegraphics[width=0.45\textwidth]{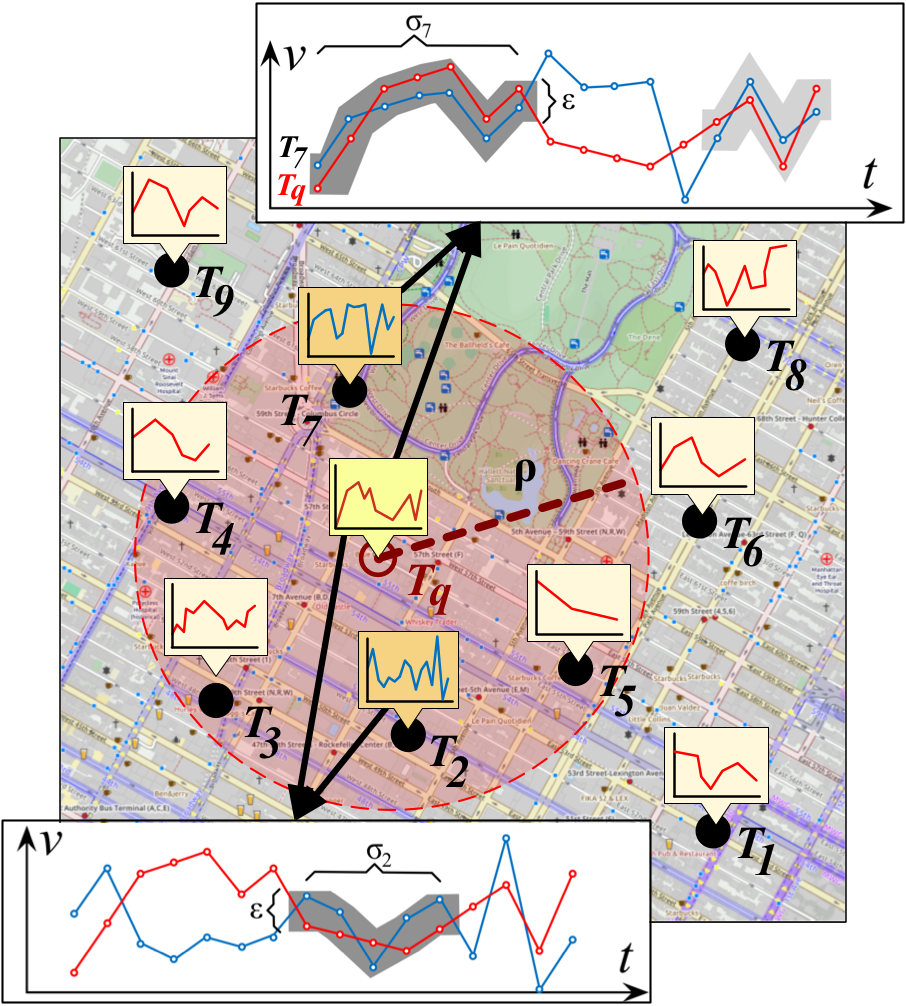}
\caption{Retrieving geolocated time series based on spatial distance and local similarity.}
\label{fig:example_query}
\end{figure}

Combining this local similarity constraint with a filter on {\em spatial distance} leads to a novel set of hybrid queries. Figure~\ref{fig:example_query} shows an example with a query time series $T_q$ searching over a set of time series $T_1,\dots,T_9$ for those within radius $\rho$ from its location and also locally similar to $T_q$. In particular, with respect to a given $\epsilon$, results should also be locally similar to $T_q$ for at least 5 consecutive timestamps. Qualifying results include $T_2$ with local similarity score $\sigma_2 = 5$ (bottom chart), and $T_7$ with $\sigma_7 = 7$ (top chart).



It turns out that such hybrid queries involving local similarity can still be evaluated using the \btsr index. We first present a baseline method employing a sweep-line algorithm to check for local similarity, and then describe how this can be optimized by using appropriately placed {\em checkpoints}, based on the local similarity score threshold specified by the query, in order to skip unnecessary comparisons. Despite the fact that this saves some computations, the resulting time savings are relatively small, since the number of index nodes that need to be probed is not essentially reduced. To overcome this problem, we introduce an improvement to the \btsr index, which is based on temporally segmenting the time series bounds within each node and deriving tighter bounds per segment. Once the time series bounds in each node become more fine-grained, pruning the search space for local similarity queries proves much more effective.

Summarizing, our main contributions are as follows:

\begin{itemize}
 \item We extend our previous work on hybrid queries for geolocated time series to support local time series similarity. We consider both range and top-$k$ queries, including combined criteria for spatial distance and local time series distance.
 \item We present how such queries can be answered efficiently exploiting the previously introduced \btsr index.
 \item To achieve greater savings in execution time by further reducing node accesses, we propose an enhanced variant of \btsr, called \sbtsr, which additionally employs temporal segmentation in each node to derive tighter, more fine-grained time series bounds.
 \item We experimentally evaluate our methods using real-world datasets from different application domains, showing that \btsr can efficiently handle hybrid queries under local similarity search, while \sbtsr achieves even higher performance due to the additional temporal segmentation.
\end{itemize}

The remainder of the paper is structured as follows. Section \ref{sec:related} reviews related work. Section \ref{sec:problem} formally defines the problem. Section \ref{sec:ls_q_btsr} presents how query evaluation under local time series similarity can be executed using the \btsr, while Section \ref{sec:sbtsr_index} presents the enhanced \sbtsr. Section \ref{sec:exp} reports our experimental results, and Section \ref{sec:conclusions} concludes the paper.

\section{Related Work}
\label{sec:related}


Similarity search over time series has provided a wide range of algorithmic approaches; a detailed survey with experimental evaluation is available in~\cite{DBLP:journals/pvldb/EchihabiZPB18}. Initially, the focus was mostly on wavelet-based methods~\cite{chan1999icde} to reduce the dimensionality of time series and generate an index based on the transformed sequences. In contrast, state-of-the-art approaches for time series indexing are based on the {\em Symbolic Aggregate Approximation} (SAX) representation \cite{jessica2007dmkd}. The first index in this family was $i$SAX\cite{shieh2008kdd}, offering multi-resolution representations for time series. Further extensions like $i$SAX 2.0~\cite{camerra2010icdm}, $i$SAX2+~\cite{camerra2014kais}, ADS+~\cite{zoumpatianos2014sigmod}, Coconut~\cite{DBLP:journals/pvldb/KondylakisDZP18}, DPiSAX~\cite{dpisaxjournal}, and ParIS~\cite{DBLP:conf/bigdataconf/PengFP18} provided a wide range of advanced capabilities. These indices support {\em global} similarity search, i.e., the similarity score is computed over the entire length of the compared time series, as opposed to {\em local} similarity, which allows to consider similar subsequences. The most recent addition to this $SAX$-based family is \textit{ULISSE}~\cite{linardi2018scalable}, which can answer similarity search queries of {\em varying} length. However, this still differs from our setting, since in \textit{ULISSE} the goal is to build an index that supports similarity search for queries of any length within a given range $[\ell_{min}, \ell_{max}]$. Furthermore, none of the aforementioned approaches supports geolocated time series, and thus cannot efficiently process hybrid queries combining conditions on spatial distance and time series similarity.


The problem of {\em subsequence matching} over time series is to identify matches of a (relatively short) query subsequence across one or more (relatively long) time series. The UCR suite \cite{rakthanmanon2012searching} offers a framework comprising various optimizations regarding subsequence similarity search. Matrix Profile~\cite{yeh2016matrix} includes methods for detecting, for each subsequence of a time series, its \textit{nearest neighbor} subsequence, by keeping track of Euclidean distances among candidate pairs. Applying such approaches in our setting is not straightforward. First, they involve Euclidean or DTW distances, which are different from our definition of local similarity score, hence the pruning heuristics do not hold in our case. Second, they do not consider geolocated time series, thus spatial filtering has to be carried out independently, which reduces pruning opportunities.


To the best of our knowledge, the only index that supports searching over geolocated time series is the \btsr~\cite{chatzig17btsr,DBLP:conf/gis/Chatzigeorgakidis18}. This hybrid index follows a similar rationale set by {\em spatio-textual indices} \cite{chen2013pvldb} that can facilitate evaluation of queries combining location-based predicates with keyword search. 
In a similar spirit, \btsr is a spatial-first index based on the R-tree that can additionally compute bounds on similarity of time series instead of a textual similarity between documents. Apart from an MBR, each node also stores bounds over the time series indexed in its subtree. Thus, it offers increased pruning capabilities for range and top-$k$ queries involving both time series similarity and spatial proximity. In the current work, we show how \btsr can be used for another family of hybrid queries involving {\em local similarity} of time series. Furthermore, we introduce a variant structure, called \sbtsr, which constructs tighter bounds over temporally segmented time series to offer stronger pruning power.
\section{Local Similarity Search on Geolocated Time Series}
\label{sec:problem}

Next, we briefly present some background on geolocated time series and the \btsr index, and then formally define the problem. 


\subsection{Preliminaries}
\label{subsec:preliminaries}

\noindent \textbf{\emph{Geolocated Time Series}}. A {\em time series} is a time-ordered sequence of values $T = \{T^1, T^2 \ldots, T^{n}\}$, where $T^{i}$ is the value at the $i$-th timestamp and $n$ is the length of the series. A geolocated time series is additionally characterized by a \emph{location}, denoted by $T.loc$. The {\em spatial distance} $d$ between two geolocated time series is the Euclidean distance of their respective locations.



\begin{figure}[tb]
    \centering
    \includegraphics[width=0.75\textwidth]{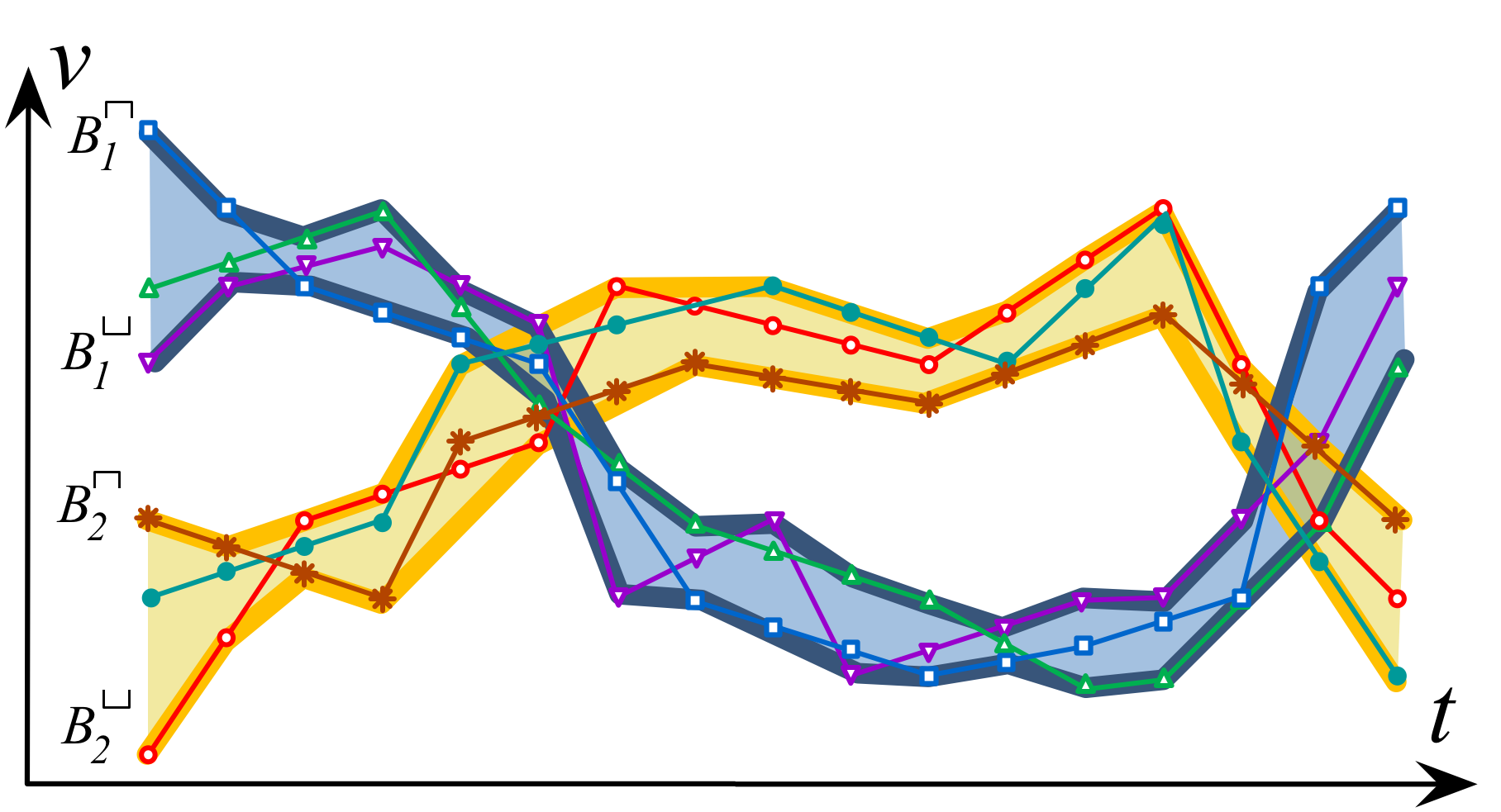}
    \caption{MBTS constructed for two sets of time series.}
    \label{fig:example_bundle}
\end{figure}

\noindent \textbf{\emph{The \btsr Index}}. In \cite{chatzig17btsr}, we have introduced the \btsr index, which is based on the notion of {\em Minimum Bounding Time Series} (MBTS). In a similar manner that an MBR encloses a set of geometries, an MBTS encloses {\em a set of time series} $\mathcal{T}$ using a pair of bounds that fully contain all of them. Figure~\ref{fig:example_bundle} depicts an example of two MBTSs for two disjoint sets of time series. Formally, given a set of time series $\mathcal{T}$, its MBTS consists of an \emph{upper bounding time series} $B^{\sqcap}$ and a \emph{lower bounding time series} $B^{\sqcup}$, constructed by respectively selecting the maximum and minimum of values at each timestamp $i \in \{ 1, \dots, n\}$ among all time series in set $\mathcal{T}$ as follows:

\begin{align}\label{eq:bounds1}
 \begin{split}
  & B^{\sqcap} = \{ \max_{T \in \mathcal{T}} T^1, \ldots, \max_{T \in \mathcal{T}} T^n \} \\
  & B^{\sqcup} = \{ \min_{T \in \mathcal{T}} T^1, \ldots, \min_{T \in \mathcal{T}} T^n \}
 \end{split}
\end{align}

\begin{figure}[!t]
 \centering
 \subfloat[Sample dataset with MBRs over locations]{\includegraphics[width=0.3\textwidth]{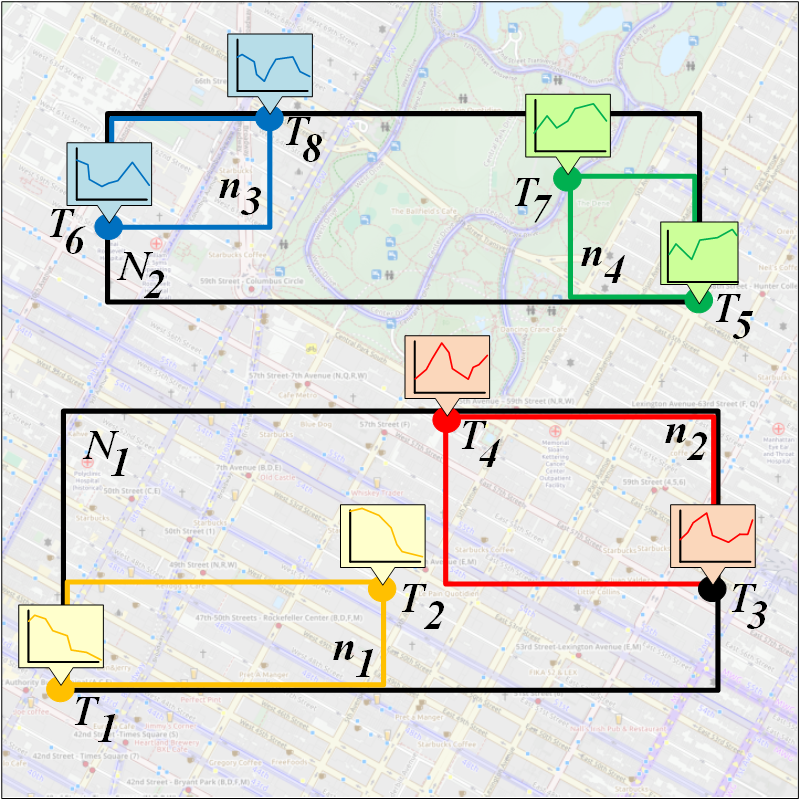}\label{subfig:sample}}
 \hspace{7pt}
 \subfloat[Spatial-only R-tree index]{\includegraphics[width=0.3\textwidth]{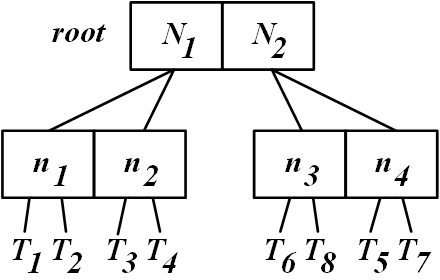}\label{subfig:rtree}}
 \hspace{7pt}
 \subfloat[Hybrid \btsr index]{\includegraphics[width=0.3\textwidth]{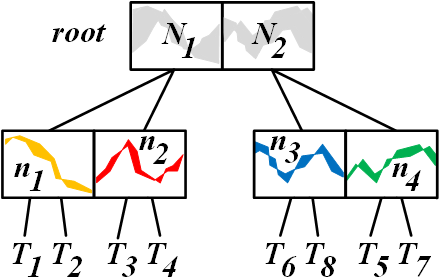}\label{subfig:btsrtree}}
 \caption{The \btsr index.}
\label{fig:example}
\end{figure}

A \btsr index is initialized as an R-tree \cite{Guttman1984} built on the spatial attributes of the given geolocated time series dataset, as depicted in the example of Figure~\ref{fig:example}. Besides MBRs, each node is enhanced to also store MBTSs, shown as colored strips per node in Figure~\ref{subfig:btsrtree}. This enables efficient pruning of the search space when evaluating hybrid queries combining time series similarity with spatial proximity.
For each child, a node stores a pre-specified number of MBTSs. Each MBTS is calculated according to Eq.~\ref{eq:bounds1}. Construction and maintenance of the \btsr follow the procedures of the R-tree for data insertion, deletion and node splitting. Objects (i.e., geolocated time series) are inserted into leaf nodes and any resulting changes are propagated upwards. Once the nodes have been populated, the MBTS of each node are calculated bottom-up, relying on $k$-\textit{means clustering} according to their Euclidean distance in the time series domain. The example in Figure \ref{fig:example_bundle} depicts the $k=2$ MBTSs (as two bands with a thick outline) obtained for a set of time series (shown as thin polylines). In a \btsr, each parent node receives all the MBTSs of its children and computes its own $k$ MBTSs. The process continues upwards, until reaching the root.

\subsection{Problem Definition}
\label{subsec:prob_def}

We first define the local similarity between time series, and then present the query variants we consider in this paper.



\begin{mydefinition}[Local Time Series Similarity]
The {\em local similarity score} $\sigma$ between two time series $T$ and $T'$ is the maximum count of consecutive timestamps during which the respective values of $T$ and $T'$ do not differ by more than a given margin $\epsilon$, i.e., $\sigma(T, T', \epsilon) = |I_{max}|$, where $I_{max}$ is the longest consecutive time interval $I$ such that $\forall i \in I, |T^{i} - T'^{i}| \leq \epsilon$.
\end{mydefinition}




In this work, our goal is to efficiently support hybrid queries on geolocated time series that retrieve the results based both on spatial proximity and local similarity. Specifically, we focus on the following types of queries (hereafter referred to as \textit{LS-queries}):

\begin{itemize}
    \item $Q_{rr}(T_q, \rho, \epsilon, \delta)$: Given a geolocated time series $T_q$, retrieve every geolocated time series $T$ such that $T$ is located within range $\rho$ from $T_q$, i.e., $d(T_q, T) \leq \rho$ and has local similarity to $T_q$ at least $\delta$, i.e., $\sigma(T_q, T) \geq \delta$.
    \item $Q_{kr}(T_q, k, \epsilon, \delta)$: Given a geolocated time series $T_q$, retrieve the spatial $k$-nearest neighbors to $T_q$ that also have local similarity to $T_q$ at least $\delta$.
    \item $Q_{rk}(T_q, \rho, \epsilon, k)$: Given a geolocated time series $T_q$, retrieve the top-$k$ geolocated time series that have the highest local similarity to $T_q$ with respect to $\epsilon$ and are located within range $\rho$ from $T_q$.
\end{itemize}

\begin{myexample}
 Figure \ref{fig:example_query} depicts an example of the $Q_{rr}(T_q, \rho, \epsilon, \delta)$ query. Given the geolocated time series $T_q$ as query, we seek the spatially close ones (i.e., within a circle of radius $\rho$) that are also locally similar within margin $\epsilon$ for at least $\delta$ timestamps. In this example, despite five geolocated time series being within range, only $T_2$ and $T_7$ qualify for the final result, since these are the ones that are also locally similar for at least one time interval of length at least $\delta$.
\end{myexample}
\section{LS-Queries Using the \btsr}
\label{sec:ls_q_btsr}
A straightforward approach for answering LS-queries would be to use a spatial index to first filter by spatial distance and then perform a sequential scan across each result to filter out those having local similarity score below the given threshold. This suffers from generating an unnecessarily large number of intermediate results which are then discarded. Instead, we propose to process LS-queries by leveraging the \btsr index \cite{chatzig17btsr}, which can prune the search space simultaneously according to both criteria. 


While traversing the \btsr, {\em spatial filtering} is performed at each node $N$ by computing the {\em bounding distance} $mindist_{sp}$ between the location of $T_q$ and the MBR of $N$, as in R-Trees~\cite{DBLP:conf/sigmod/RoussopoulosKV95}.

\begin{figure}[tb]
    \centering
    \includegraphics[width=0.75\textwidth]{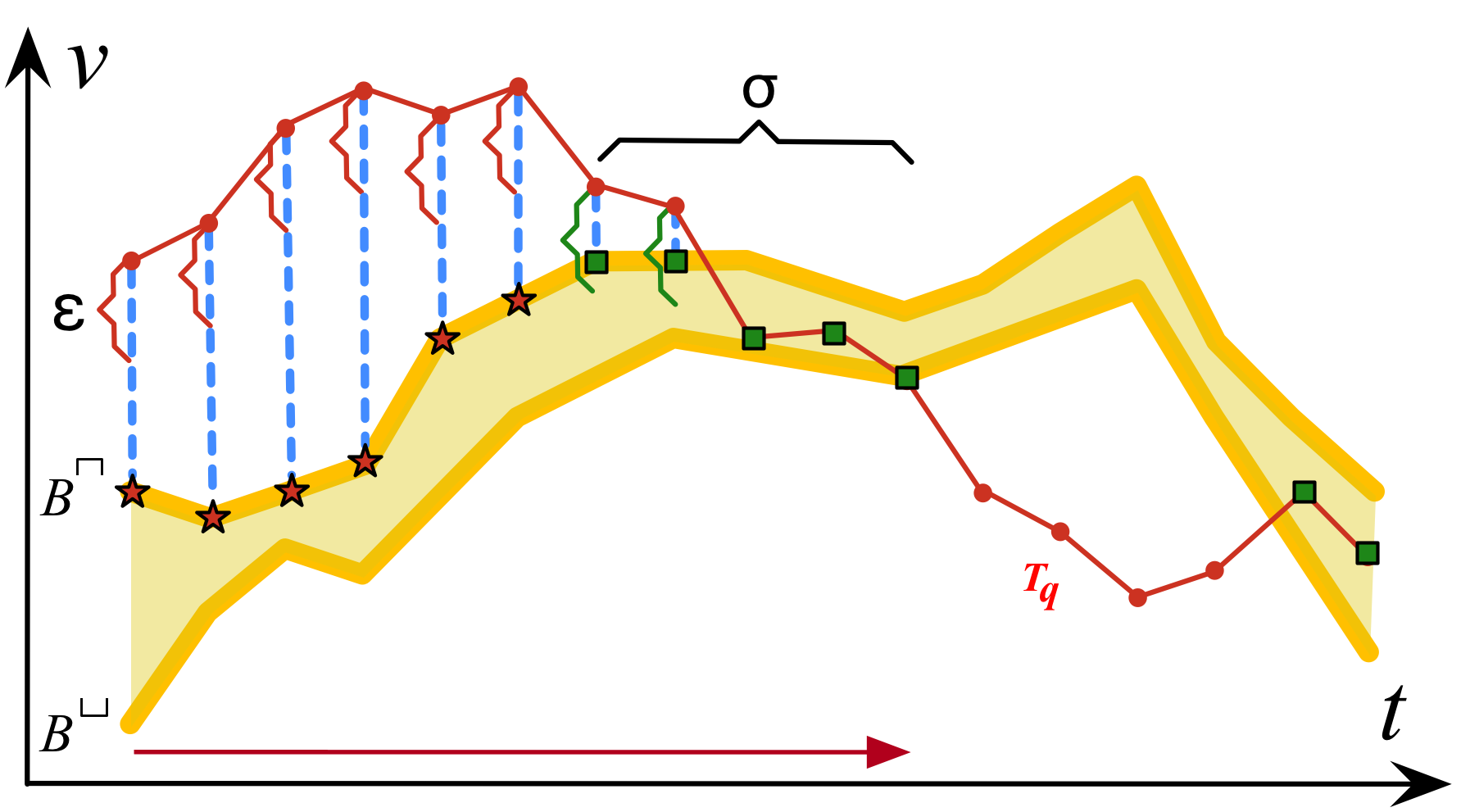}
    \caption{Local similarity check against an MBTS.}
    \label{fig:sim_mbts}
\end{figure}

For {\em time series similarity}, we exploit the MBTS stored within each node. Considering an MBTS at a node $N$, we calculate its distance $mindist_{ts}^i$ from $T_q$ at each timestamp $i$ as:
  
\begin{equation}
 \begin{split}
  mindist_{ts}^i(T_q, {MBTS}_N) = \begin{cases}
	T_q^i - B_{N}^{\sqcap_{i}}, & \text{if} \;\; T_q^i > B_{N}^{\sqcap_{i}} \\
	B_{N}^{\sqcup_{i}} - T_q^i, & \text{if} \;\; T_q^i < B_{N}^{\sqcup_{i}} \\
	0, & \text{if} \;\; B_{N}^{\sqcap_{i}} \leq T_q^i \leq B_{N}^{\sqcup_{i}}
	  \end{cases}
 \end{split}
 \label{eq:mindist_ts}
\end{equation}

\noindent where $B_{N}^{\sqcap_{i}}$ and $B_{N}^{\sqcup_{i}}$ are the upper and lower values of the MBTS at timestamp $i$. By definition of MBTS, no time series indexed under $N$ can differ from $T_q$ by less than $mindist_{ts}^i$ at timestamp $i$. Hence, only at those timestamps that $mindist_{ts}^i \leq \epsilon$, it is possible that a time series indexed under $N$ is locally similar to $T_q$. Subsequently, we can compute a {\em local similarity bound} $\sigma_B$:
\begin{equation}
\sigma_{B}(T_q, MBTS_N, \epsilon) = max\{|I|; \forall i \in I, mindist^i_{ts}(T_q, MBTS_N) \leq \epsilon\}.
\label{eq:sim_bound}
\end{equation}
\noindent that reflects the {\em maximum} interval $I$ of consecutive timestamps where the distance computed by Eq.~\ref{eq:mindist_ts} does not exceed margin $\epsilon$. 
This value is an upper bound of the local similarity scores of $T_q$ with any time series enclosed in this MBTS. Figure \ref{fig:sim_mbts} shows that $T_q$ deviates from the given MBTS by no more than $\epsilon$ during two intervals: one consisting of $|I_1|=5$ consecutive timestamps and a smaller one with only $|I_2|=2$ timestamps (shown as square points). So, the local similarity bound for this MBTS is $\sigma_{B}=5$.

By construction, the MBTSs of a child node $N'$ get tighter bounds compared to those of its parent $N$ as we descend the \btsr. It is easy to verify that 
\begin{equation}
\begin{split}
\sigma_{B}(T_q, MBTS_N, \epsilon) \geq \sigma_{B}(T_q, MBTS_{N'}, \epsilon)
\end{split}
\label{eq:maxdur_ts}
\end{equation}
\noindent hence local similarity bounds can only diminish when descending the index. This bound provides a useful pruning condition during search with a cutoff threshold $\delta$. Any node where all its MBTSs have local similarity bound $\sigma_{B}$ below $\delta$ can be safely pruned.

Next, we describe a baseline approach that employs a sequential scan over MBTSs, and then we present an optimization that prioritizes selected \textit{checkpoints} to avoid many point-wise comparisons.

\subsection{Sweep Line Approach}
\label{sec:baseline}

We explain how the \btsr can be used, in conjunction with a simple sweep-line algorithm, to answer each of the three LS-queries, taking advantage of the two types of bounds, $mindist_{sp}$ and $mindist_{ts}$, described above.


\vspace{2mm}

\noindent $\mathbold{Q_{rr}(T_q, \rho, \epsilon, \delta)}$: We traverse the \btsr starting from its root. At each inner node $N$, we first check whether $mindist_{sp}(T_q, MBR_N)$ $\leq$ $\rho$. If so, we employ a sweep line across the time axis to compute the local similarity bound $\sigma_{B}(T_q, MBTS_N, \epsilon)$ for every MBTS included in $N$. If {\em all} resulting bounds $\sigma_{B}$ are below $\delta$, the subtree under $N$ is pruned. Otherwise, the search continues at the children. Upon reaching a leaf node, we fetch the geolocated time series contained therein, and verify the query constraints against each one. Each $T$ such that $d(T_q.loc, T.loc) \leq \rho$ and $\sigma(T_q, T, \epsilon) \geq \delta$ is added to the results.

\vspace{2mm}

\noindent $\mathbold{Q_{kr}(T_q, k, \epsilon, \delta)}$: We maintain a priority queue $P$ containing both inner nodes (sorted by ascending $mindist_{sp}$) and geolocated time series (sorted by ascending spatial distance to $T_q$). We start by adding to $P$ the root of \btsr. In each iteration, we retrieve the top element from $P$. If it is an inner node, we visit its children to calculate local similarity bounds $\sigma_B$ according to Eq.~\ref{eq:sim_bound}. For any child $N$ that $\sigma_B$ of one of its MBTSs satisfies threshold $\delta$, we search the subtree of $N$. Then, we calculate the corresponding spatial distance ($mindist_{sp}$ for a node $N$ or Euclidean distance for a geolocated time series $T$) and insert it back to $P$. Once we encounter a geolocated time series $T$ at the top of $P$, we add it to the results. The process terminates once $k$ geolocated time series have been obtained.

\vspace{2mm}

\noindent $\mathbold{Q_{rk}(T_q, \rho, \epsilon, k)}$: This query is evaluated similarly to the previous one, with two differences. The first difference is that the priority queue $P$ is now sorted based on local similarity bounds in descending order, instead of spatial distance bounds in ascending order. The second is that before inserting an item (node or time series) to $P$, its spatial distance ($mindist_{sp}$ or exact) is calculated, and if it is higher than $\rho$ the item is skipped. The traversal starts again from the root, and terminates once $k$ time series have been retrieved from the top of $P$. These are the top-$k$ results with respect to local similarity (if another time series $T$ had higher local similarity, it would have been retrieved from $P$ first), and they are located within range $\rho$ from $T_q$ (otherwise, they would not have been admitted to $P$).


\subsection{Checkpoint Approach}
\label{sec:checkpoint}


The drawback of the sweep-line approach is that it needs to perform a comparison for each individual timestamp to eventually determine the exact or maximum local similarity of a given time series or node, respectively. In the following, we explain how we can use \textit{checkpoints} along the time axis to avoid this exhaustive search. These checkpoints prioritize specific timestamps when checking for candidate matches to eagerly filter out non-qualifying items.


Assume a query with local similarity threshold $\delta$. We can place checkpoints at every $\delta$ timestamps, and only apply the local similarity filter (i.e., $|T_q^{i} - T^{i}| \leq \epsilon$) at those. If no checkpoint satisfies the condition, this item can be safely pruned since it cannot have local similarity to $T_q$ at least $\delta$ (as this would require the condition to be true for at least $\delta$ consecutive timestamps, thus crossing at least one checkpoint).

\begin{figure}[!t]
 \centering
 \subfloat[Checkpoints placed every $\delta$ timestamps.]{\includegraphics[width=0.45\textwidth]{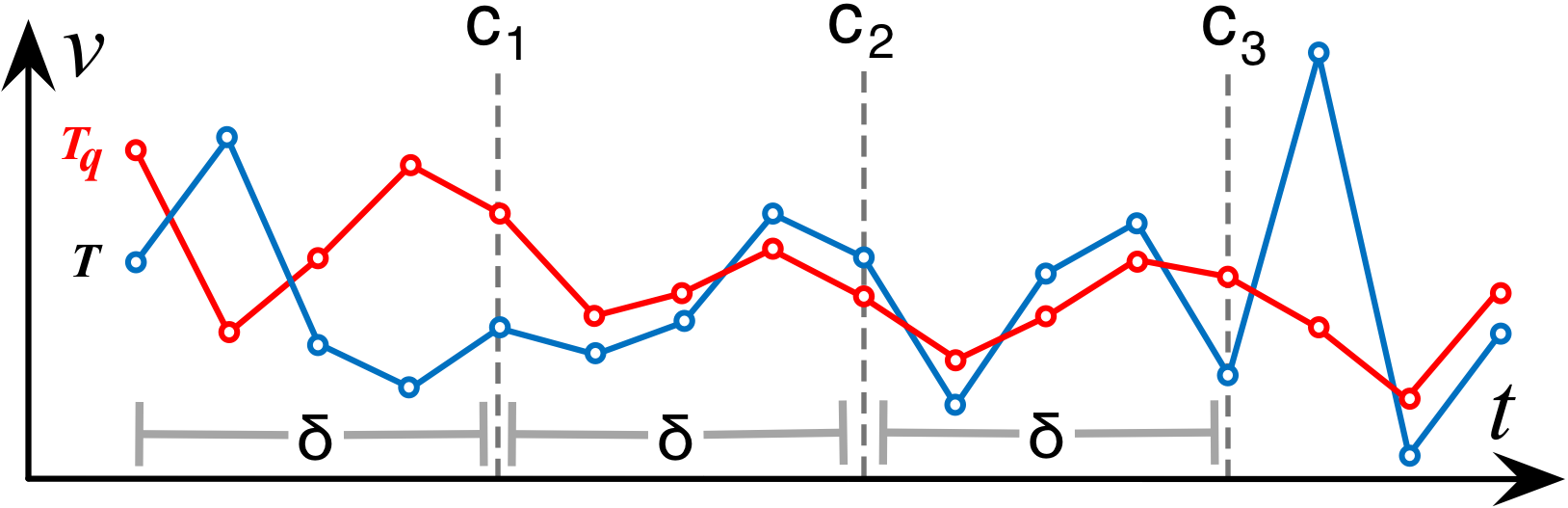}\label{fig:checkpoints}}
 \hspace{5pt}
 \subfloat[Local similarity starting before checkpoint at $t'$.]{\includegraphics[width=0.45\textwidth]{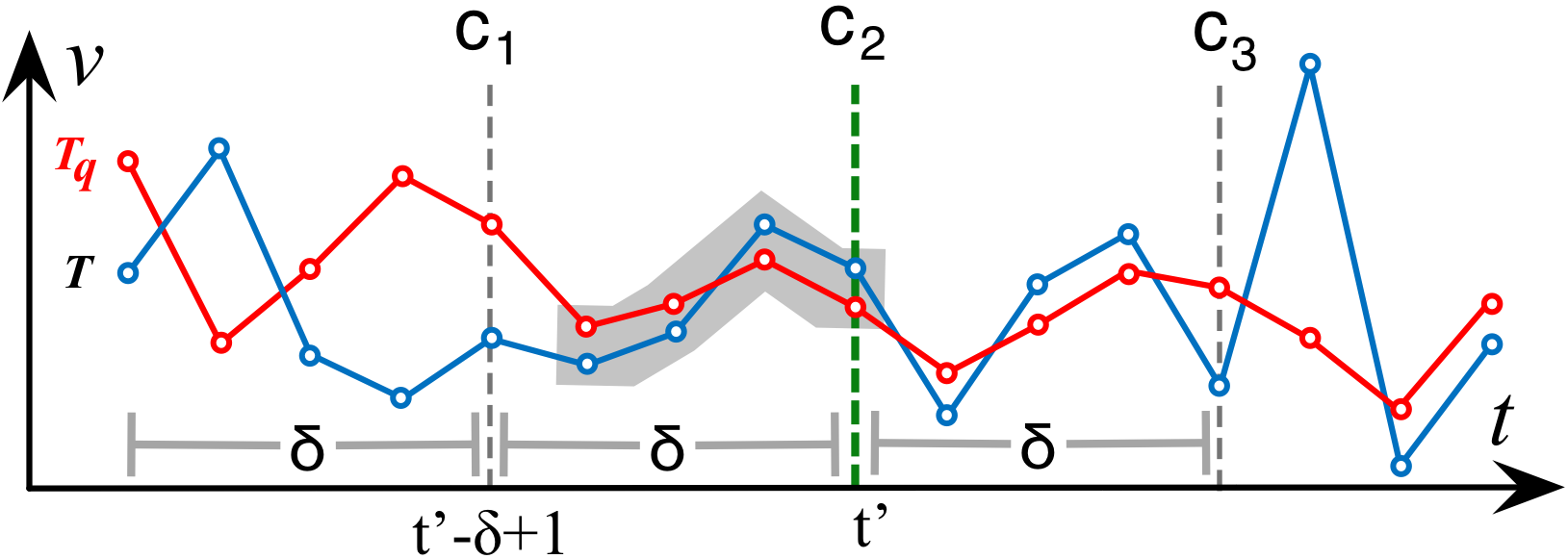}\label{fig:proof1}} \\
 \subfloat[Local similarity ending after checkpoint at $t'$.]{\includegraphics[width=0.45\textwidth]{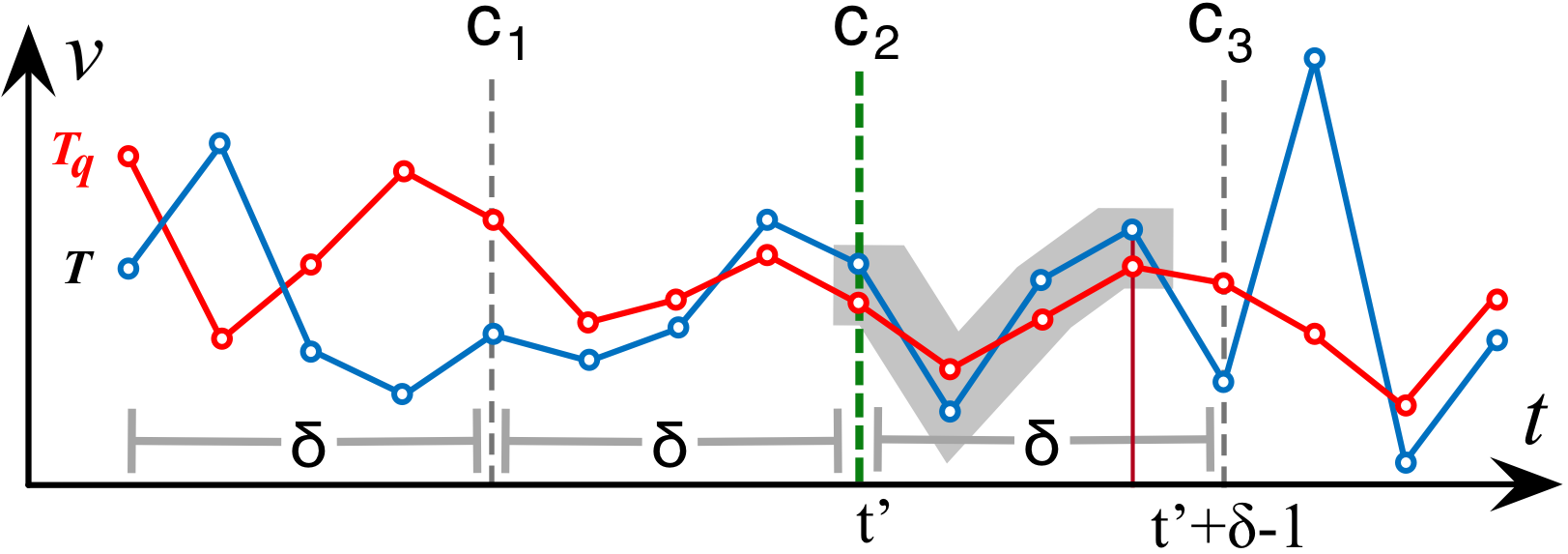}\label{fig:proof2}}
 \caption{Local time series similarity via checkpoints.}
 \label{fig:proof}
\end{figure}

Figure \ref{fig:checkpoints} shows an example with checkpoints placed along the time axis every $\delta=5$ timestamps. For clarity, we consider a single time series $T$. Assume a checkpoint at timestamp $t'$ and a minimal duration $\delta$ starting at timestamp $t'-\delta+1$ for asserting local similarity with query $T_q$, as shown with the grey strip in Figure \ref{fig:proof1}. This interval cannot have smaller duration, as it would not satisfy the $\delta$ constraint. Thus, the local similarity condition will evaluate to true at checkpoint $t'$. Similarly, if such an interval ends at timestamp $t'+\delta-1$ (Figure \ref{fig:proof2}), it will be detected at the checkpoint at $t'$. This observation entails that it suffices to check for local similarity only at checkpoints, i.e., every $\delta$ timestamps. We denote the set of checkpoints as $C$, determined at query time. If a checkpoint satisfies the condition, then we need to scan both forward and backward from it to determine the actual local similarity score, i.e., to find the exact extent of the time interval for which the condition holds.

\begin{figure}[tb]
    \centering
    \includegraphics[width=0.75\textwidth]{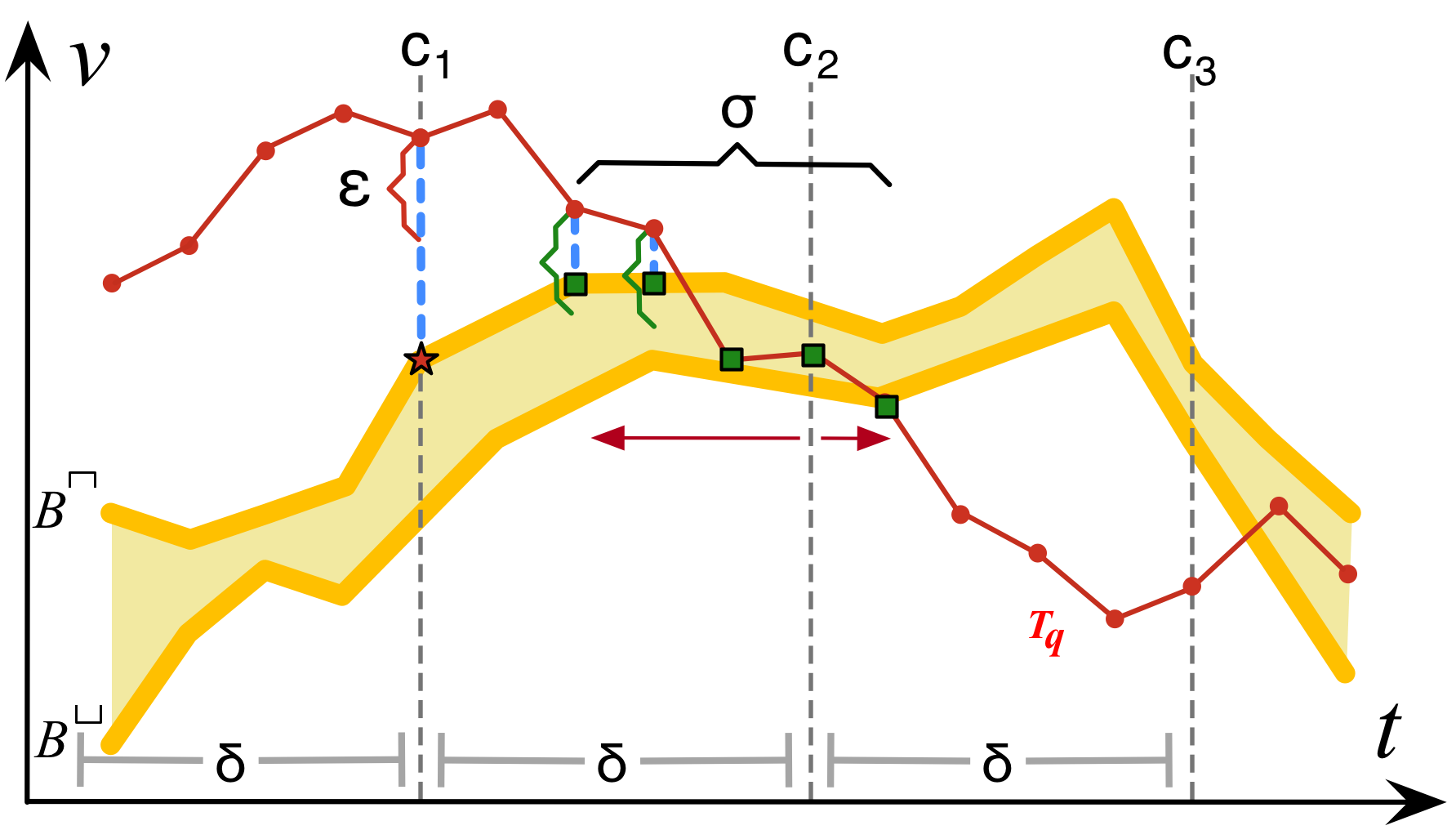}
    \caption{Local similarity with a MBTS using checkpoints.}
    \label{fig:sim_mbts_checkpoints}
\end{figure}

Figure \ref{fig:sim_mbts_checkpoints} exemplifies the use of checkpoints for comparing $T_q$ to an MBTS of a node for $\delta=5$ timestamps. Instead of sequentially performing 11 comparisons until verifying that local similarity score $\sigma$ is at least $\delta$ (i.e., we stop the verification at $t=11$, once $\sigma=5$), we check only around the checkpoints. At the leftmost checkpoint $c_1$, no local similarity is found ($T_q$ is farther than $\epsilon$ from the MBTS), so we skip directly to checkpoint $c_2$. Since $T_q$ differs by less than $\epsilon$ at $c_2$, we need to compare values backward and forward, up to the previous and next checkpoint, respectively. This requires only 6 comparisons instead of 11 to decide that this node may contain candidates. Next, we describe how probing with checkpoints is applied during evaluation of LS-queries.

\vspace{2mm}

\noindent $\mathbold{Q_{rr}(T_q, \rho, \epsilon, \delta)}$: Algorithm \ref{alg:query_r} outlines the procedure. Initially, we obtain the children of the root node in a list and place the checkpoints every $\delta$ timestamps (Lines 1-3). We iterate over each item $N$ in this list. If $N$ is an inner node, we have to examine whether both constraints with respect to $\rho$ and $\delta$ are met for each of its children. Verification of MBTS against query $T_q$ will be discussed shortly. If this is the case, we traverse the sub-tree of each child in the same manner, by adding it to the list (Lines 7-11), thus descending the tree. If the examined node is a leaf (Line 12), we iterate over each contained time series $T$ to check the constraints $\rho$ and $\delta$. If $T$ qualifies, it is added to the results (Lines 13-15). Note that now the calculation of local similarity scores for geolocated time series is based on checkpoints (Line 14), as discussed above.

Verification of MBTS against the local similarity constraints $\epsilon, \delta$ is applied using checkpoints (Lines~17-38). This verification concerns each MBTS in a given node $N'$. At each checkpoint $c$, we first verify whether its $mindist_{ts}^c$ to query $T_q$ is at most $\epsilon$ (Line 20). If so, we first scan backward to inspect whether there are at least $\delta$ consecutive timestamps where $T_q$ deviates by at most $\epsilon$ from this MBTS (Lines 22-29). Similarly, we probe forward from checkpoint $c$ (Lines 30-37). In either case, once local similarity no longer holds at a timestamp, probing skips to the next checkpoint. If the check fails for all checkpoints of all MBTSs, then this node cannot contain any results (Line 38).

\SetAlFnt{\small}
\begin{algorithm}[tb]
	\DontPrintSemicolon
	\vspace{4pt}
	$R \leftarrow \emptyset$ \\
	$List \leftarrow Root.entries$ \\
	$C \leftarrow determineCheckpoints(\delta)$ \\
	\While{$List \neq \emptyset$}{
		$N \leftarrow List.getNext()$ \\
		\If{$N$ is not leaf}{
			\ForEach{$N' \in N.getChildren()$}{
				\If{$mindist_{sp}(T_q, MBR_{N'}) \leq \rho$}{
				    $count \leftarrow \emptyset$ \\
                    \If{$VerifyMBTS(T_q, N', C, \epsilon, \delta)$}{
                        $List \leftarrow List \cup \{N'.getChildren()\}$ \\
                    }
				}
			}
		}
		\Else{
			\ForEach{$T \in N.getObjects()$}{
				\If{$d(T_q, T) \leq \rho \land \sigma^C(T_q, T, \epsilon) \geq \delta$}{
					$R \leftarrow R \cup \{ T \}$ \\
				}
			}
			\KwRet R
		}
	}
	
	\vspace{4pt}
    \SetKwProg{verifyMBTS}{Procedure}{}{}
    \verifyMBTS{$VerifyMBTS(T_q, N', C, \epsilon, \delta)$}{
        \ForEach{$MBTS \in N'$}{
    	    \ForEach{$c \in C$}{
                \If{$mindist_{ts}^c(T_q, MBTS) \leq \epsilon$}{
                    $count++, c' \leftarrow c$ \\
                    \While{True}{
                        $c'--$ \\
                        \If{$mindist_{ts}^{c'}(T_q, MBTS) \leq \epsilon$}{
                            $count++$ \\
                            \If{$count \geq \delta$}{
                                \KwRet True
                            }
                        }
                        \Else{
                            $break$
                        }
                    }
                    \While{True}{
                        $c++$ \\
                        \If{$mindist_{ts}^c(T_q, MBTS) \leq \epsilon$}{
                            $count++$ \\
                            \If{$count \geq \delta$}{
                                \KwRet True
                            }
                        }
                        \Else{
                            $break$
                        }
                    }
                } 
            }
        }
        \KwRet $False$
	}

	\caption{$Q_{rr}(T_q, \rho, \epsilon, \delta)$}
	\label{alg:query_r}
\end{algorithm}

\vspace{2mm}

\noindent $\mathbold{Q_{kr}(T_q, k, \epsilon, \delta)}$: We follow a similar procedure to the one in Section~\ref{sec:baseline} for query $Q_{kr}$, employing the same verification process over MBTSs and time series as in Algorithm~\ref{alg:query_r}. Algorithm~\ref{alg:query_k} describes the procedure. We start by adding the root node to a priority queue $P$ based on spatial distance (Line 2). After determining the checkpoints using the given $\delta$ (Line 3), we iteratively retrieve elements from $P$ (Line 5). Then, three cases may occur: 
\begin{enumerate}
\item[(i)] If this element is a time series (Lines 6-9), it is guarranteed to be a result, given that $P$ is sorted based on spatial distance from $T_q$. Indeed, any subsequent element must be located farther than the current. When list $R$ obtains the required number $k$ of results, the search terminates. 
\item[(ii)] The element is a leaf node (Lines 10-14): In this case, we obtain each time series $T$ contained in this leaf, and verify the local similarity score of $T$ against $\delta$. If the condition is met, we calculate the spatial distance of candidate $T$ from query $T_q$ and push $T$ into the priority list along with its spatial distance (Lines 10-14). 
\item[(iii)] If the element is an inner node, we iterate over its children and only push back to the queue the ones whose MBTSs are verified against $\epsilon$ and $\delta$ using checkpoints (Lines 15-19).
\end{enumerate}

\SetAlFnt{\small}
\begin{algorithm}[!t]
	\DontPrintSemicolon
	$R \leftarrow \emptyset$ \\
	$P.push(Root)$ \\
	$C \leftarrow determineCheckpoints(\delta)$ \\
	\While{$P$ is not empty}{
		$N \leftarrow P.poll()$ \\
		\If{$N$ is raw}{
				$R \leftarrow R \cup \{ N \}$ \\
			\If{$|R| = k$}{
				$break$ \\
			}
		}
		\ElseIf{$N$ is leaf}{
			\ForEach{$T \in N.getObjects()$}{
				\If{$\sigma^C(T_q, T, \epsilon) \geq \delta$}{
					 $T.dist \leftarrow d(T_q, T)$ \\
					 $P.push(T, T.dist)$ \\					 
				}					
			}
		}
		\Else{
			\ForEach{$N' \in N.getChildren()$}{
				\If{$VerifyMBTS(T_q, N', C, \epsilon, \delta)$}{
					$N'.dist \leftarrow mindist_{sp}(T_q, MBR_{N'})$ \\
					$P.push(N', N'.dist)$ \\	
				}
			}
		}
	}
	\KwRet R
	\caption{$Q_{kr}(T_q, k, \epsilon, \delta)$}
	\label{alg:query_k}	
\end{algorithm}

\vspace{2mm}

\noindent $\mathbold{Q_{rk}(T_q, \rho, \epsilon, k)}$: The procedure for this query is listed in Algorithm~\ref{alg:query_ks}. Notice that for employing checkpoints, we need a local similarity threshold $\delta$, so as to determine their placement, but this query does not specify a fixed $\delta$. To be able to obtain one during search, we now maintain two priority queues: $P$ holds inner nodes sorted by local similarity bounds (Eq.~\ref{eq:sim_bound}), while $R$ keeps up to $k$ geolocated time series sorted by local similarity scores (as in Def. 1). We initially set $\delta=1$, so checkpoints are trivially placed at every timestamp. This implies that computation of local similarity scores with $\delta=1$ is equivalent to the sweep line approach. However, $\delta$ increases with the detection of qualifying results, hence checkpoints will progressively get placed more sparsely. The search starts by adding the \btsr root in $P$ (Line 2). We iteratively poll the top element from $P$, and there are two possible cases:
\begin{enumerate}
\item[(i)] The top element is a leaf node. Then, we iterate over the contained time series and add the ones that satisfy the spatial condition ($\rho$) to $R$, along with their corresponding local similarity score $\sigma$ if it exceeds the current value of $\delta$ (Lines 8-12).
Once $R$ exceeds capacity $k$, its last element is evicted to make room for the newly inserted one and $\delta$ is updated according to the local similarity score $\sigma_k$ of the $k$-th element in $R$. In this case, the placement of checkpoints is re-adjusted according to the increased $\delta$ value (Lines 13-16).

\item[(ii)] The top element is an inner node. In this case, we iterate over each child $N'$ and check if $mindist_{sp}(T_q, {MBR_N'}) \leq \rho$. If $N'$ qualifies, we calculate the local similarity bound $\sigma_B$ of all its MBTSs using checkpoints. If the maximum among these bounds $max(\sigma_B) \geq \delta$, then $N'$ is inserted to $P$ with this maximum score (Lines 17-25).
\end{enumerate}

The process terminates once the top element in $P$ has local similarity less than $\delta$ (Lines 6-7). The result is the contents of $R$.

\SetAlFnt{\small}
\begin{algorithm}[!t]
	\DontPrintSemicolon
	$R \leftarrow \emptyset$ \\
	$P.push(Root)$ \\
	$\delta \leftarrow 1$ \\
	$C \leftarrow determineCheckpoints(\delta)$ \\
	\While{$P$ is not empty}{
		\If{$P.peekFirst.\sigma_B < \delta$}{
		    $break$ \\
		}
		\If{$N$ is leaf}{
			\ForEach{$T \in N.getObjects()$}{
				\If{$d(T_q, T) \leq \rho$}{
					 \If{$\sigma^C(T_q, T, \epsilon) \geq \delta$}{
					    $R.push(T, \sigma^C(T_q, T, \epsilon))$ \\					 
					 }
				}
				\If{$R.size > k$}{
				    $R.pollLast$ \\
				    $\delta \leftarrow R.peekLast.\sigma$ \\
				    $C \leftarrow determineCheckpoints(\delta)$ \\
			    }
			}
		}
		\Else{
			\ForEach{$N' \in N.getChildren()$}{
				\If{$mindist_{sp}(T_q, MBR_{N'}) \leq \rho$}{
				    $\sigma_B \leftarrow 0$ \\
				    \ForEach{$MBTS \in N'$}{
				        \If{$\sigma_B^C(T_q, MBTS, \epsilon) \geq \sigma_B$}{
				            $\sigma_B \leftarrow \sigma_B^C(T_q, MBTS, \epsilon)$
				        }
				    }
					\If{$\sigma_B \geq \delta$}{
					    $P.push(N', \sigma_B)$ \\	
					}
				}
			}
		}
	}
	\KwRet R
	\caption{$Q_{rk}(T_q, k, \rho)$}
	\label{alg:query_ks}	
\end{algorithm}

\section{The \texorpdfstring{\sbtsr} IIndex}
\label{sec:sbtsr_index}

\subsection{Index Structure}
\label{subsec:structure_sbtsr}
\begin{figure}[tb]
 \centering
 \subfloat[Example of a node's MBTS.]{\includegraphics[width=0.45\textwidth]{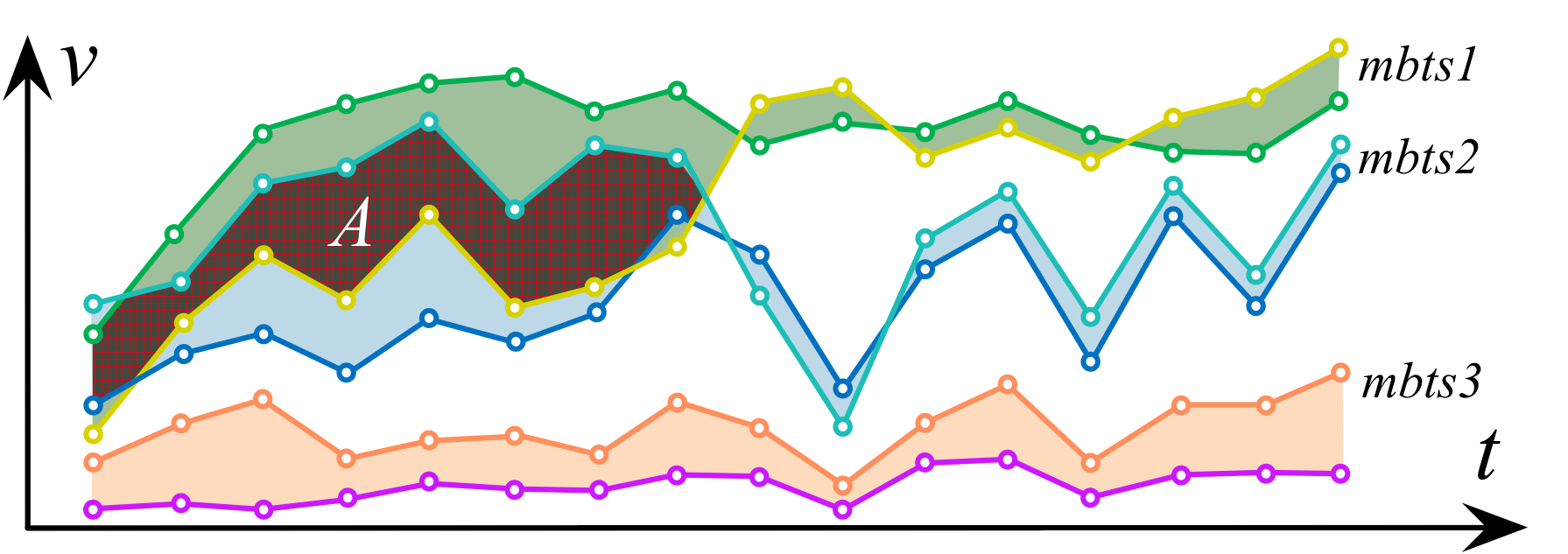}\label{subfig:bundles}}
 \subfloat[Segmenting can eliminate whitespace.]{\includegraphics[width=0.45\textwidth]{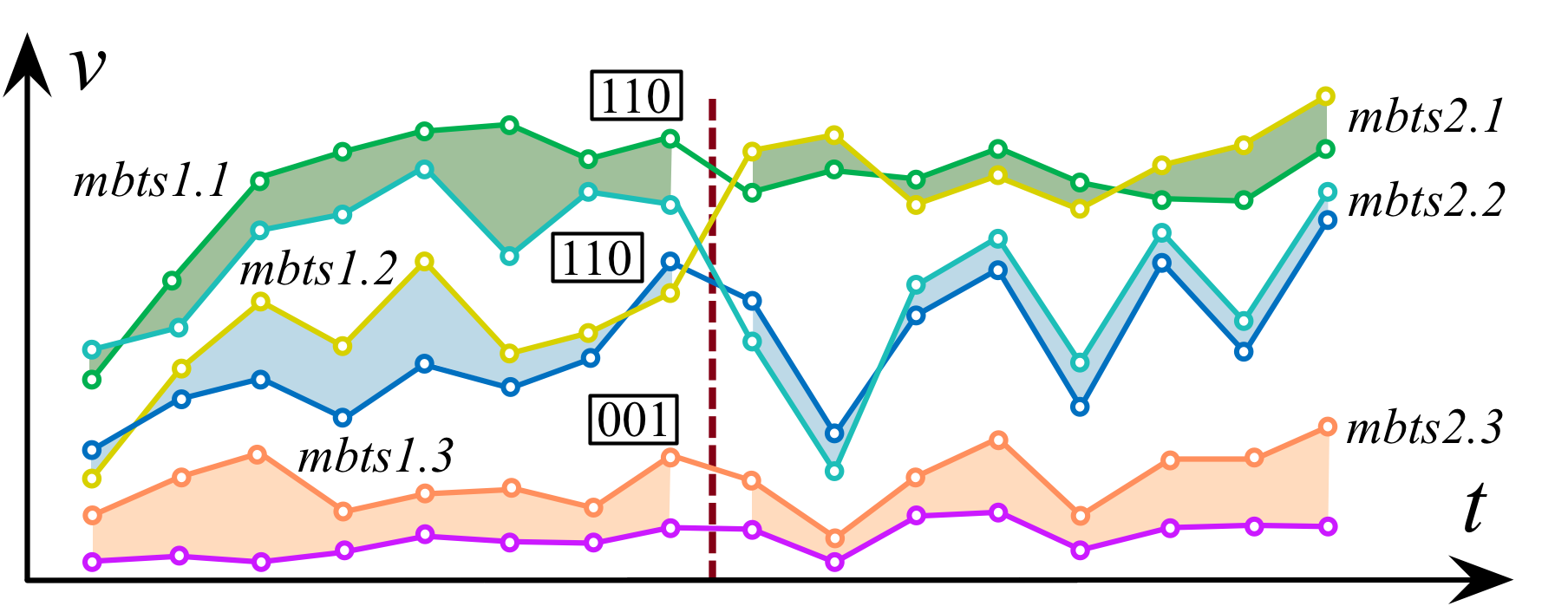}\label{subfig:bundles_part}}
 \caption{Segmenting time series yields more tight MBTS.}
 \label{fig:proof_partition}
\end{figure}

The \btsr index uses $k$-means clustering to cluster the time series under each node and then stores the MBTSs of those clusters. However, clustering entire time series typically generates many overlapping MBTSs, incurring much dead space. This has a negative impact on the pruning power of the index, especially when considering local similarities. Figure~\ref{subfig:bundles} depicts such a case of six time series indexed in a node. A $k$-means clustering with $k=3$ will form the depicted MBTSs denoted with shaded colors. As a result, the dark area $A$ represents the overlap between $mbts.1$ and $mbts.2$ and actually makes those bounds less tight. Hence, such MBTSs inflate estimates for local similarity bounds, and thus lead to unnecessarily descending further down the index.

To reduce the amount of overlap within the MBTSs of nodes, we introduce an extended version of the \btsr, named \sbtsr. \sbtsr attempts to eliminate as much overlap as possible, through segmentation of time series. Figure~\ref{subfig:bundles_part} depicts the intuition. If we segment the time series before applying $k$-means, the resulting MBTSs for each segment tend to be tighter, eliminating the excessive overlap $A$ from Figure~\ref{subfig:bundles}. The \sbtsr is built similarly to \btsr. The only difference is that the MBTSs of each node are calculated \textit{per segment}. In this method, we assume a pre-defined number $s$ of segments, but segmentation is orthogonal to our problem and can be carried out by applying existing methods like \cite{bingham2006segmentation}. Ultimately, \sbtsr allows for more aggressive pruning when traversing the index.

\subsection{Cross-Segment Continuity Via Bit-Vectors}
\label{subsec:bit_vectors}
A downside of the segmentation approach is the loss of the MBTS continuity across time, which results in MBTSs enclosing different time series in neighboring segments. For example, in Figure~\ref{subfig:bundles_part}, there are no MBTSs in the right segment containing the same time series as $mbts1.1$ and $mbts1.2$, a fact which hinders the calculation of local similarity on the segment boundaries (the vertical line). To overcome this, we introduce a \textit{bit-vector} $V$ along each MBTS of a segment, having one bit for each MBTS created. If in the current segment a bit in vector $V$ of a given MBTS is set, this indicates that this MBTS encloses at least one common time series with another MBTS$'$ in the next segment. In the example shown in Figure~\ref{subfig:bundles_part}, $V=110$ for $mbts1.1$ indicates common time series with $mbts2.1$ and $mbts2.2$ in the next segment, while $V=001 $ for $mbts1.3$ signifies common time series with only $mbts2.3$. This way, to calculate local similarity, we can easily identify all the MBTSs that share common time series among two successive segments. 

\begin{figure}[tb]
    \centering
    \includegraphics[width=0.75\textwidth]{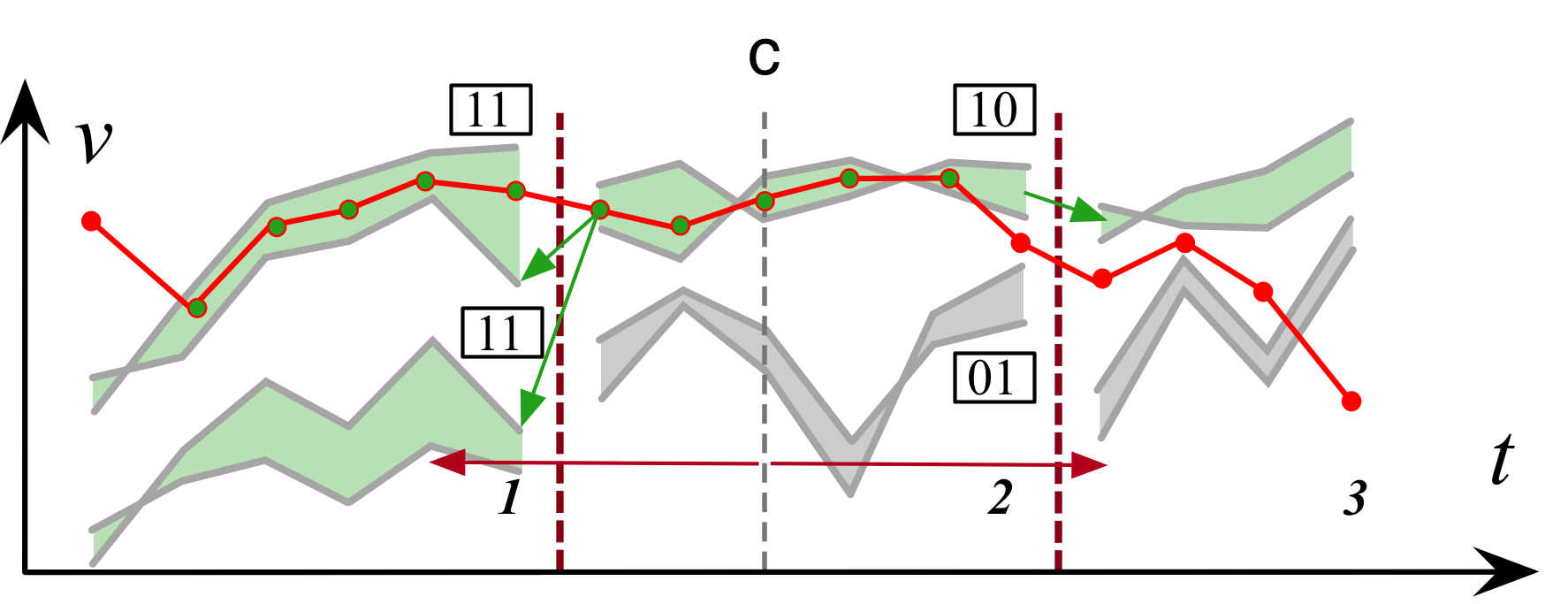}
    \caption{Example of verifying a \sbtsr node.}
    \label{fig:part_checkpnt}
\end{figure}

To evaluate LS-queries, traversal of the \sbtsr index follows a similar rationale to the procedure in Section~\ref{sec:checkpoint}. For each checkpoint $c$, we first obtain the segment where it falls in, and we scan each MBTS leftward and rightward from $c$, as discussed in Section~\ref{sec:checkpoint}. If we cross the border to another segment, the available bit-vectors directly identify the MBTS that need be examined in this neighboring segment. This propagates until the local similarity constraints ($\epsilon$ and $\delta$) are satisfied. Figure~\ref{fig:part_checkpnt} illustrates an example of a node verification. Let us consider a predetermined number of three segments and the corresponding MBTS of each segment for that node. Suppose that there exists a checkpoint $c$ on the second segment. To verify whether this node satisfies the local similarity constraints, we start from checkpoint $c$ and we check leftwards whether $mindist_{ts}^{i} \leq \epsilon$ for each timestamp. If the currently examined timestamp falls in the first segment, we fetch the corresponding MBTS and bit-vectors and continue checking whether $mindist_{ts}^{i} \leq \epsilon$ in both MBTS (green shaded), as their bit-vectors both indicate common members with the first one in segment 2. A similar procedure is followed rightwards, where we only have to check the first MBTS, according to the bit-vectors.

\section{Experimental Evaluation}
\label{sec:exp}

Next, we report results from a comprehensive evaluation of our methods against real-world datasets.

\subsection{Experimental Setup}
\label{subsec:evaluation_setup}

\vspace{2mm}

\subsubsection{Datasets}
We use three real-world datasets (Table \ref{tab:datasets}) selected from different application domains, containing diverse types of geolocated time series, as detailed below:

\begin{table}[t]
	\centering
	\caption{Datasets and parameters used in the experiments.}
	\begin{small}
	\centering
	\begin{tabular}{lccc|cccc}
	\hline
	\multirow{2}{*}{Dataset} & Area & Number of & Length of & \multicolumn{4}{c}{Default query parameters} \\
	 & (km$^2$) & locations & timeseries & $\rho$ & $\epsilon$ & $\delta$ & $k$ \\
	\hline
	Flickr & Earth & 414,967 & 96 & 30\% & 7.5\% & 20 & 30\\
    Crime & 392,000 & 362,215 & 76 & 30\% & 7.5\% & 25 & 30 \\
    Taxi & 2,500 & 417,960 & 168 & 30\% & 10\% & 20 & 30 \\
	\hline
	\end{tabular}
	\end{small}
	\label{tab:datasets}
\end{table}

\vspace{2mm}

\noindent \emph{\textbf{UK historical crime data (Crime)}}. Contains time series representing the temporal variation in the number of crime incidents reported across England and Wales over 76 months (December 2010-- March 2017). We generated time series over a grid with cell size 200 meters applied on the original data\footnote{\url{https://data.police.uk/data/}}. For each month, we counted incidents having their location within each cell.

\vspace{2mm}

\noindent \emph{\textbf{Flickr geotagged photos (Flickr)}}. Contains time series data extracted from geolocated Flickr images between 2006 and 2013 over the entire planet\footnote{\url{https://code.flickr.net/category/geo/}}. To get meaningful geolocated time series, we partitioned the space by a uniform grid of of $7200 \times 3600$ cells (each one spanning $0.05$ decimal degrees in each dimension) and counted the number of photos contained in every cell each month. We excluded empty cells (e.g., in the oceans). Each time series conveys the visits pattern (in terms of number of photos taken per month) of that region over this period.

\vspace{2mm}

\noindent \emph{\textbf{NYC taxi dropoffs (Taxi)}}. Contains time series extracted from yellow taxi rides in New York City during 2015. The original data\footnote{\url{http://www.nyc.gov/html/tlc/html/about/trip_record_data.shtml}} provide pick-up and drop-off locations, as well as corresponding timestamps for each ride. For each month, we generated time series by applying a uniform spatial grid over the entire city (cell side was 200 meters) and counting all drop-offs therein for each day of the week at the time granularity of one hour. Thus, we obtained the number of drop-offs for $24 \times 7$ time intervals in every cell, which essentially captures the weekly fluctuation of taxi destinations there. Without loss of generality, the centroid of each cell is used as the geolocation of the corresponding time series. 

\vspace{2mm}

\noindent \emph{\textbf{Synthetic}}. To test scalability, we augmented the Flickr dataset by slightly moving each location in a random manner and altering each time series value by a random number between $1$ and $10$. We produced three additional synthetic datasets each containing $\times 2$, $\times 3$, $\times 4$ the number of time series from the original dataset. 


\subsubsection{Index and Query Parameters}
To evaluate the performance benefits observed in the experiments only based on pruning, we tuned the index parameters to fixed values. The minimum ($m$) and maximum ($M$) number of entries stored in each node are set to $40$ and $100$, respectively. For both \btsr and \sbtsr, the number of MBTS to 10 and for \sbtsr, the number of segments $s$ is also set to 10. The query parameters involve the spatial distance and local similarity thresholds, i.e., $\rho$, $\epsilon$, $\delta$ and $k$. The values of these parameters are set differently for each dataset, based on their characteristics; default values are shown in Table \ref{tab:datasets}. The value of $\rho$ is set relatively, by setting the covered area as a percentage of the total area. Similarly, $\epsilon$ is set as a percentage of the maximum difference between the observed values.

\subsubsection{Evaluation Setting}
Each experiment is performed using a randomly selected workload of 100 queries for each dataset and we report the average response time. All indices are held in memory, while the leafs contain pointers to files with geolocated time series stored on disk. All methods were developed in Java. Tests were executed on a server with 4 CPUs, each containing 8 cores clocked at 2.13GHz, and 256 GB RAM running Debian Linux.

\subsection{Query Performance}
\label{subsec:query_perf}

We compare the average per query execution time for all three queries using sweep line and checkpoint methods on \btsr and the checkpoint method on \sbtsr.

\begin{figure}[htbp]
\centering
\subfloat{\fbox{\includegraphics[width=0.35\textwidth]{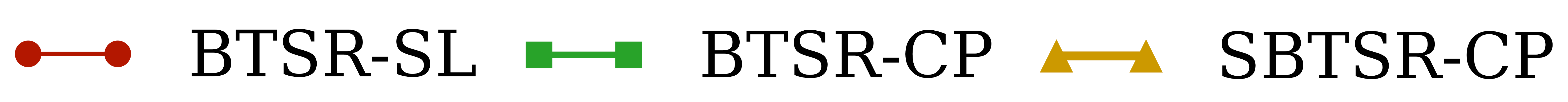}}}
\\
\vspace{-5pt}
\subfloat[Crime]{\includegraphics[trim=0.5cm 0.5cm 0.5cm 0.5cm, clip, width=0.3\textwidth]{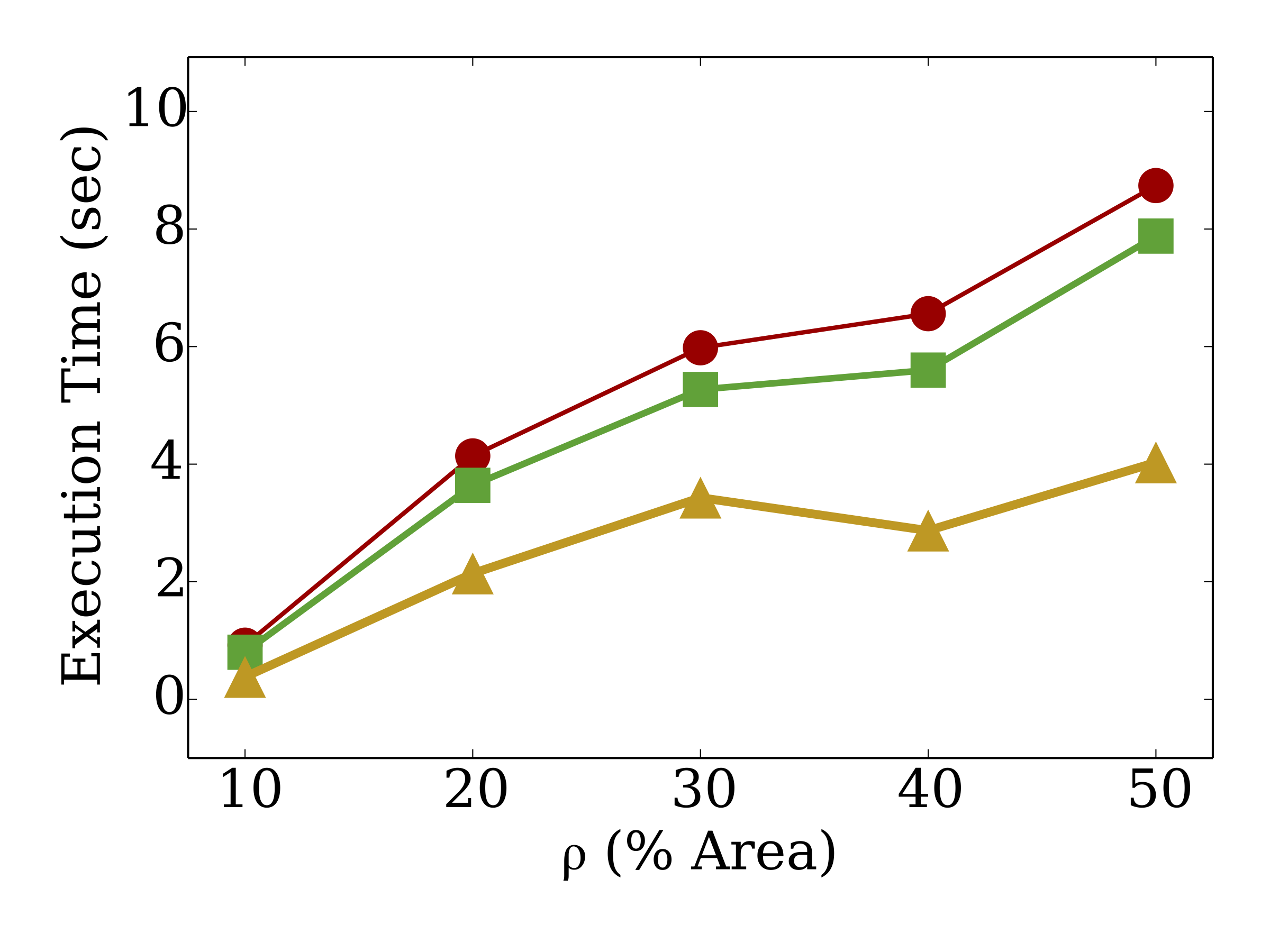}\label{subfig:var_epsSP_crime}} \quad
\subfloat[Crime]{\includegraphics[trim=0.5cm 0.5cm 0.5cm 0.5cm, clip, width=0.3\textwidth]{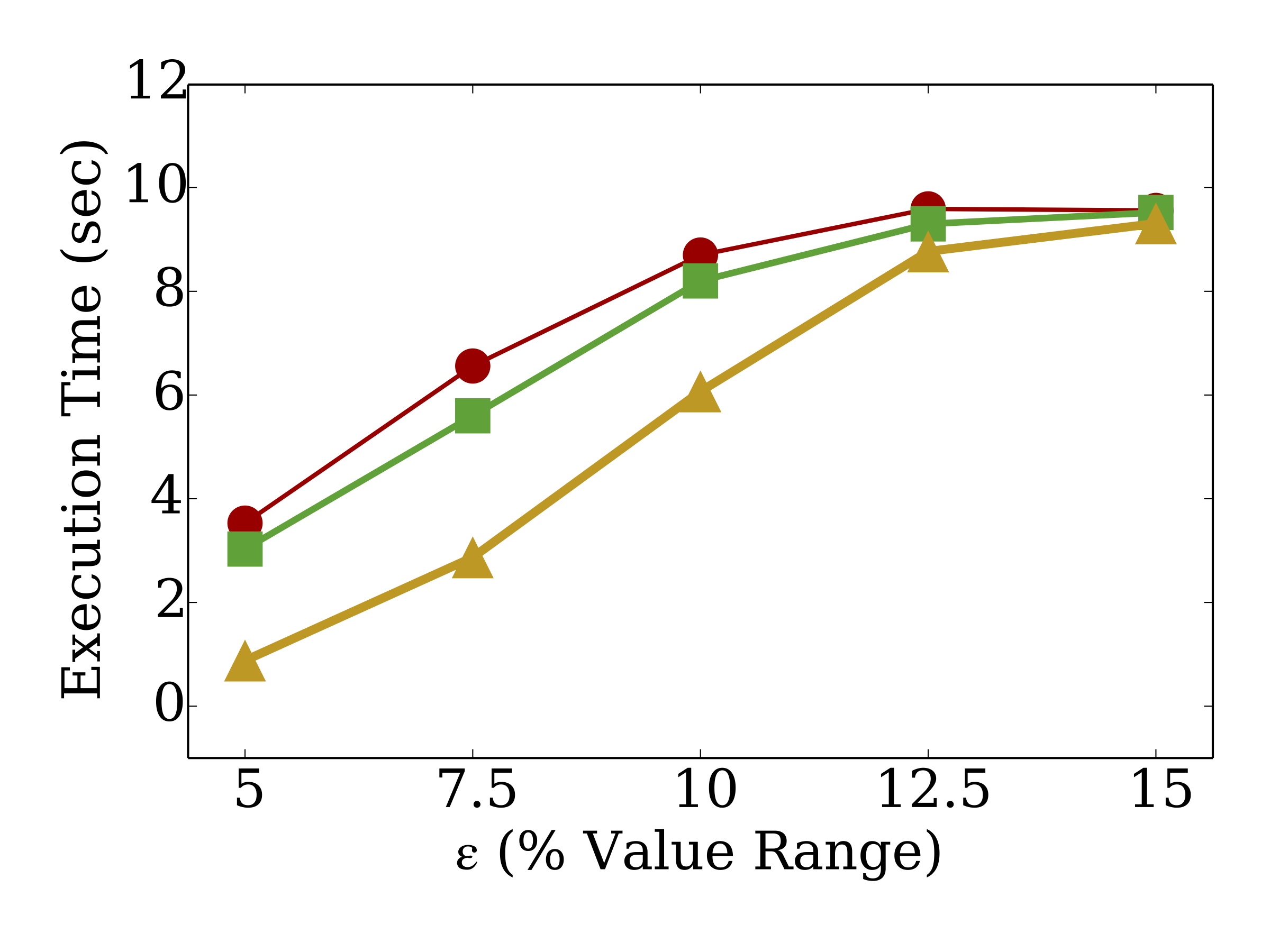}\label{subfig:var_epsTS_crime}} \\
\vspace{-5pt}
\subfloat[Flickr]{\includegraphics[trim=0.5cm 0.5cm 0.5cm 0.5cm, clip, width=0.3\textwidth]{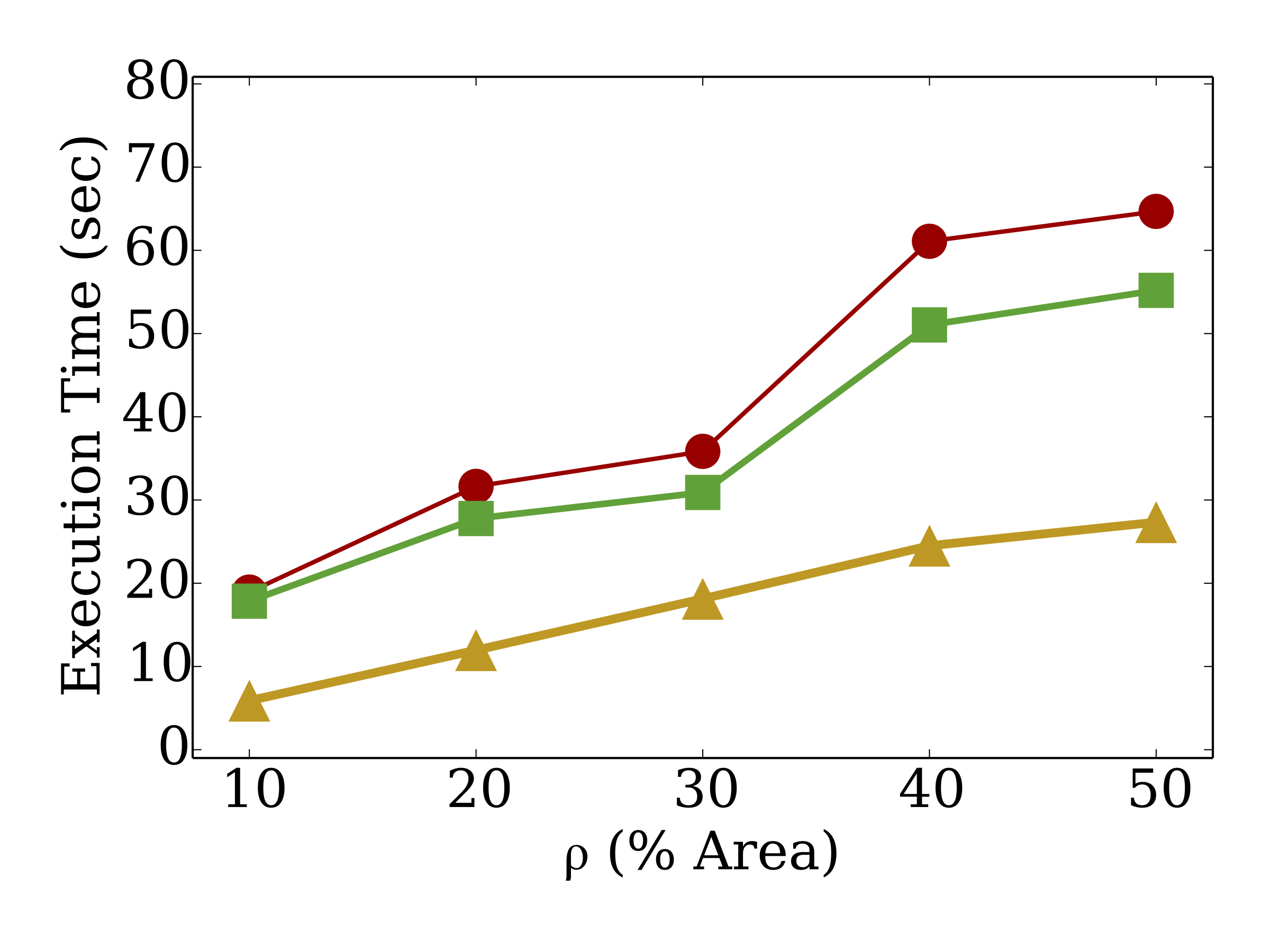}\label{subfig:var_epsSP_flickr}} \quad
\subfloat[Flickr]{\includegraphics[trim=0.5cm 0.5cm 0.5cm 0.5cm, clip, width=0.3\textwidth]{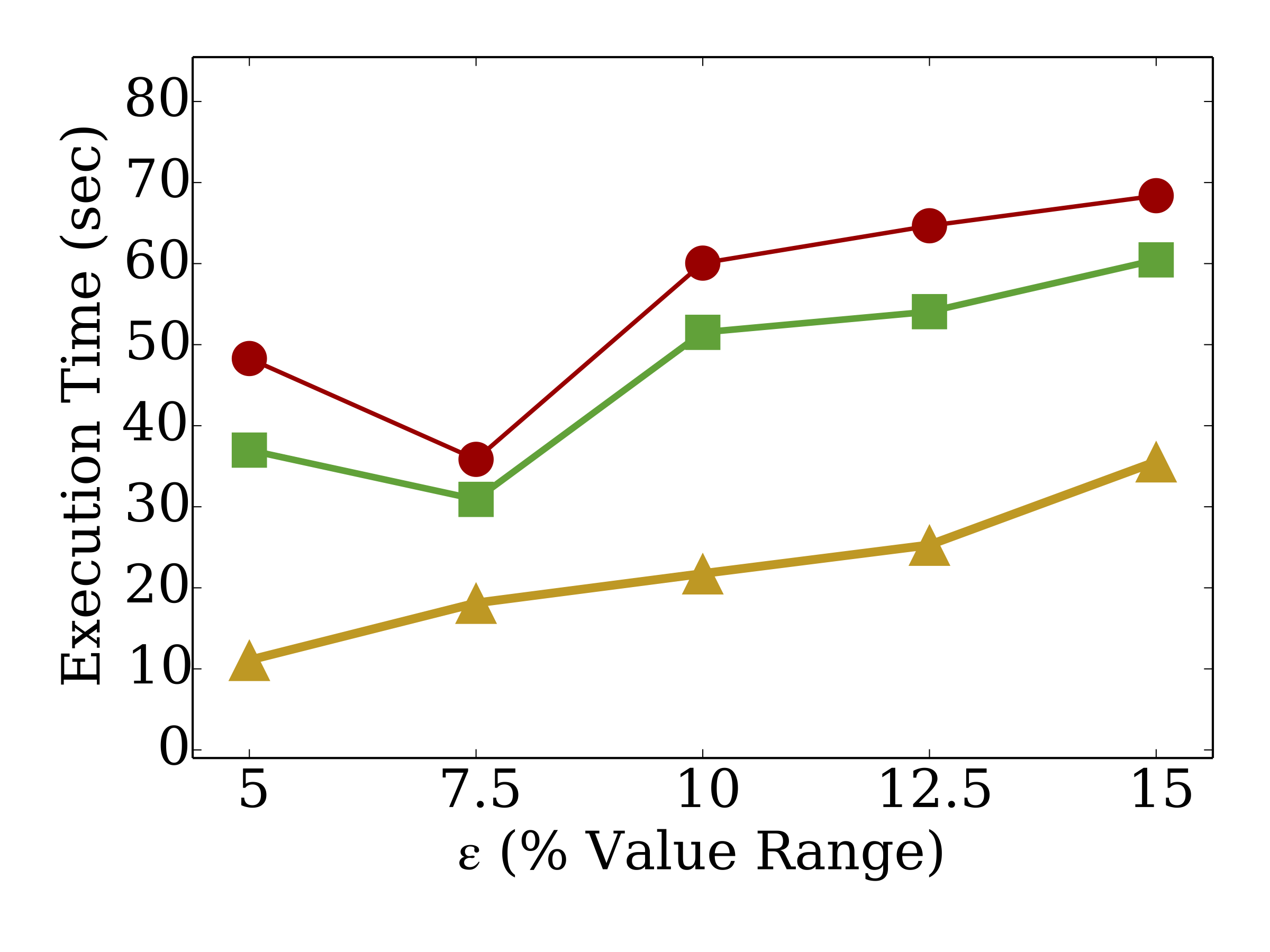}\label{subfig:var_epsTS_flickr}} \\
\vspace{-5pt}
\subfloat[Taxi]{\includegraphics[trim=0.5cm 0.5cm 0.5cm 0.5cm, clip, width=0.3\textwidth]{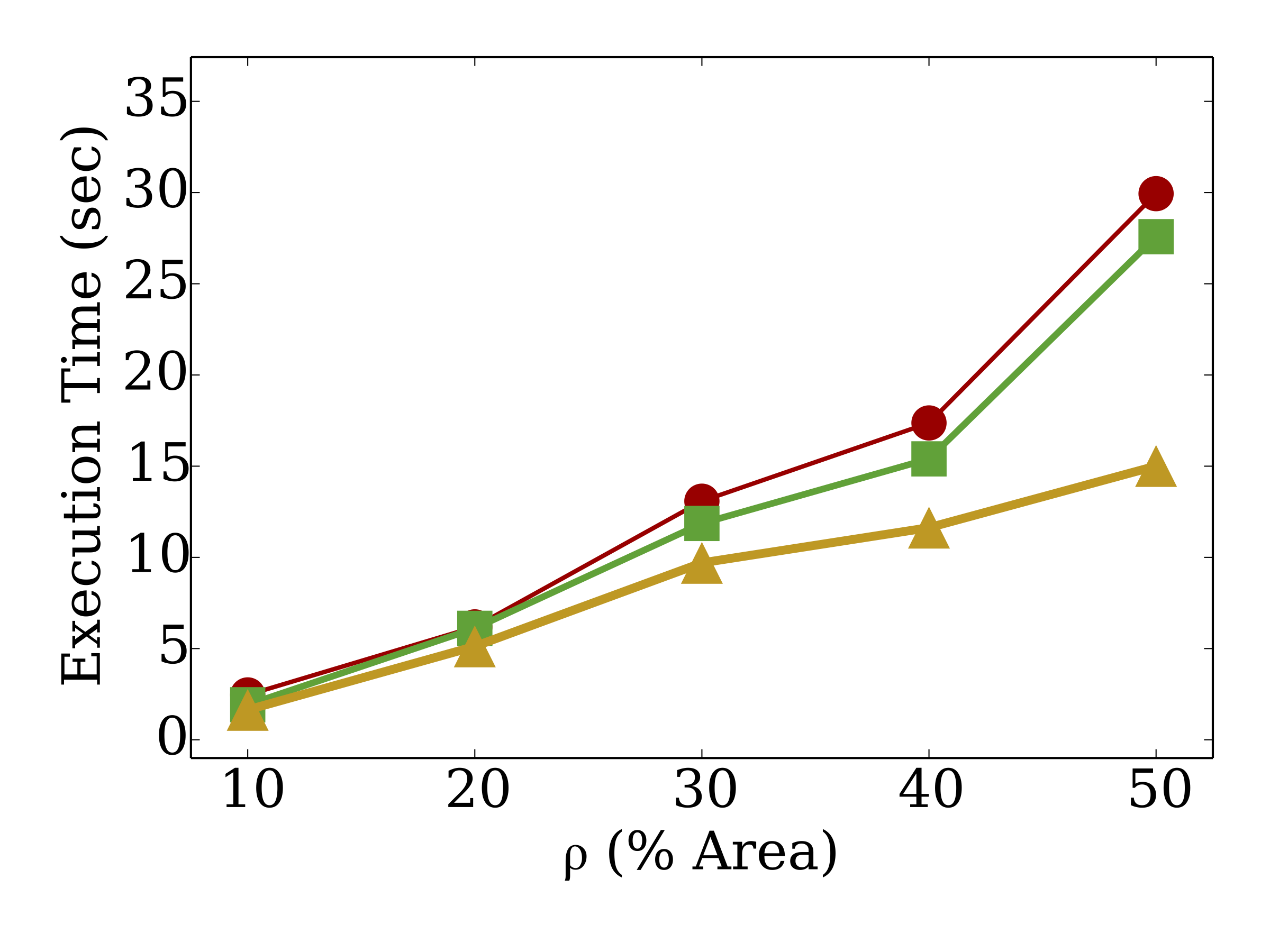}\label{subfig:var_epsSP_taxi}} \quad
\subfloat[Taxi]{\includegraphics[trim=0.5cm 0.5cm 0.5cm 0.5cm, clip, width=0.3\textwidth]{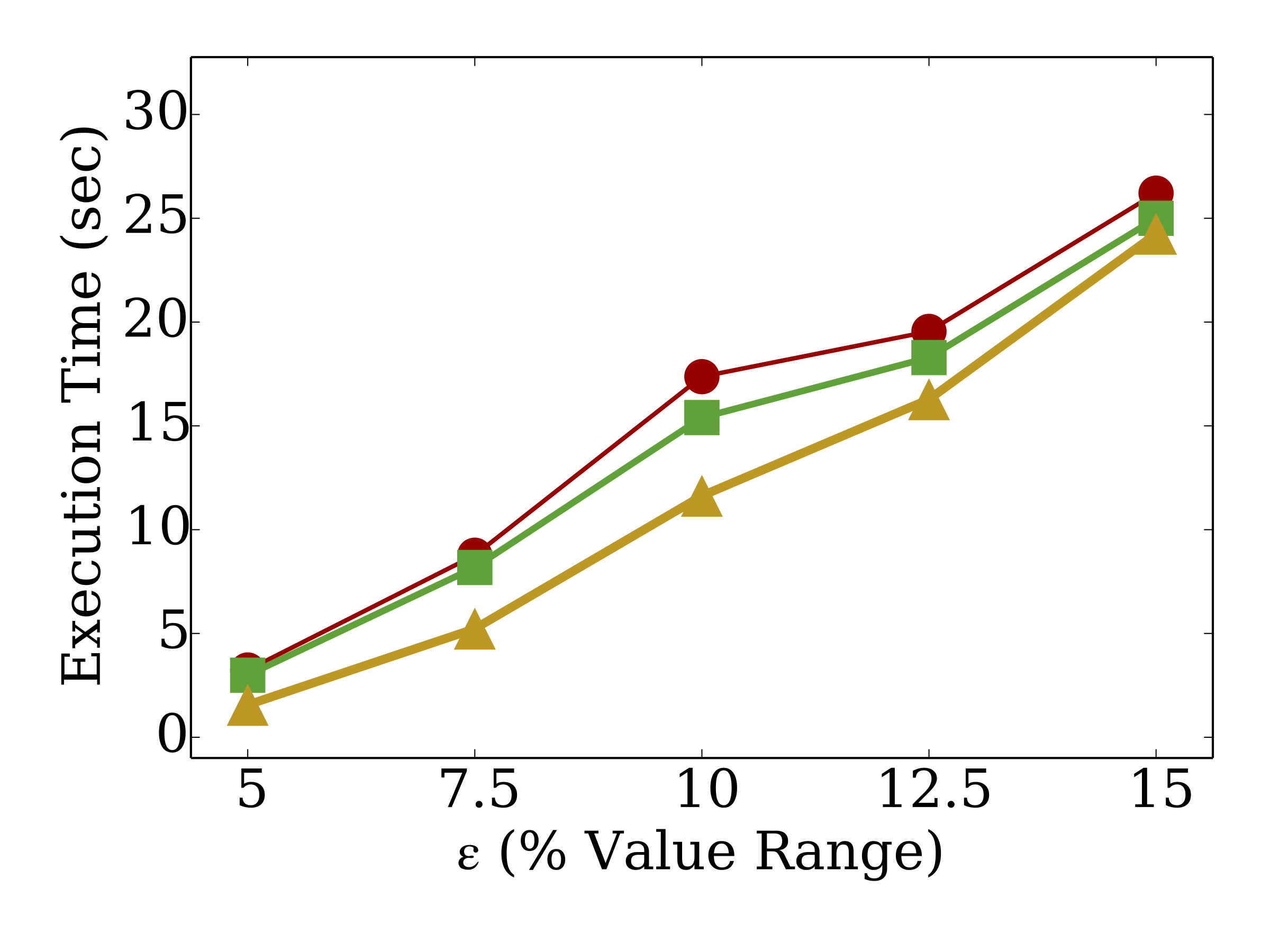}\label{subfig:var_epsTS_taxi}}
\caption{Query $Q_{rr}(T_q, \rho, \epsilon, \delta)$ for varying $\rho$ and $\epsilon$.}
\label{fig:query1a}
\end{figure}

\begin{figure*}[htbp]
\centering
\subfloat{\fbox{\includegraphics[width=0.35\textwidth]{Figures/legend.png}}}
\\
\vspace{-5pt}
\subfloat[Crime ($Q_{rr}$)]{\includegraphics[trim=0.5cm 0.5cm 0.5cm 0.5cm, clip, width=0.225\textwidth]{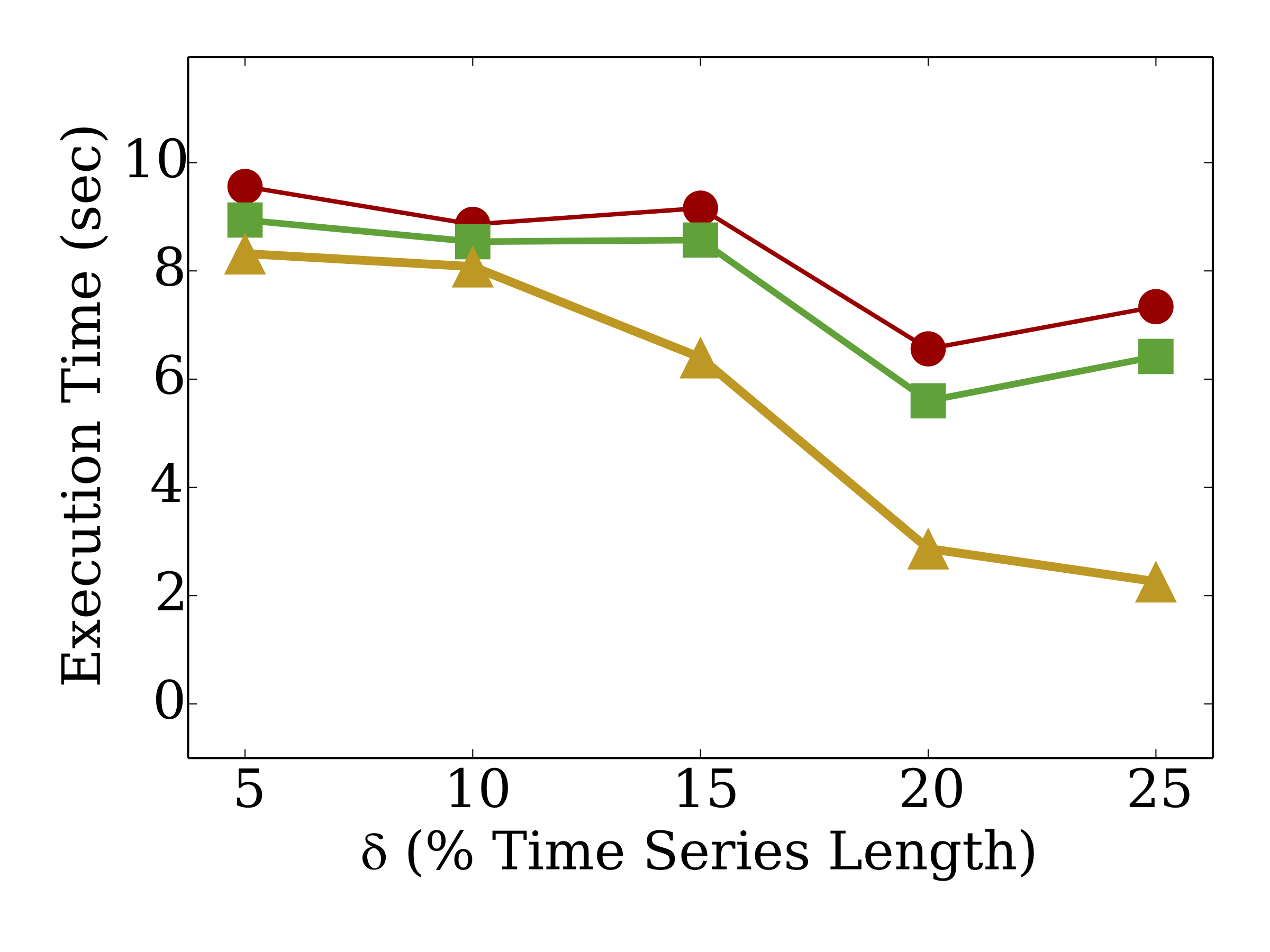}\label{subfig:var_delta_crime}}
\subfloat[Crime ($Q_{kr}$)]{\includegraphics[trim=0.5cm 0.5cm 0.5cm 0.5cm, clip, width=0.225\textwidth]{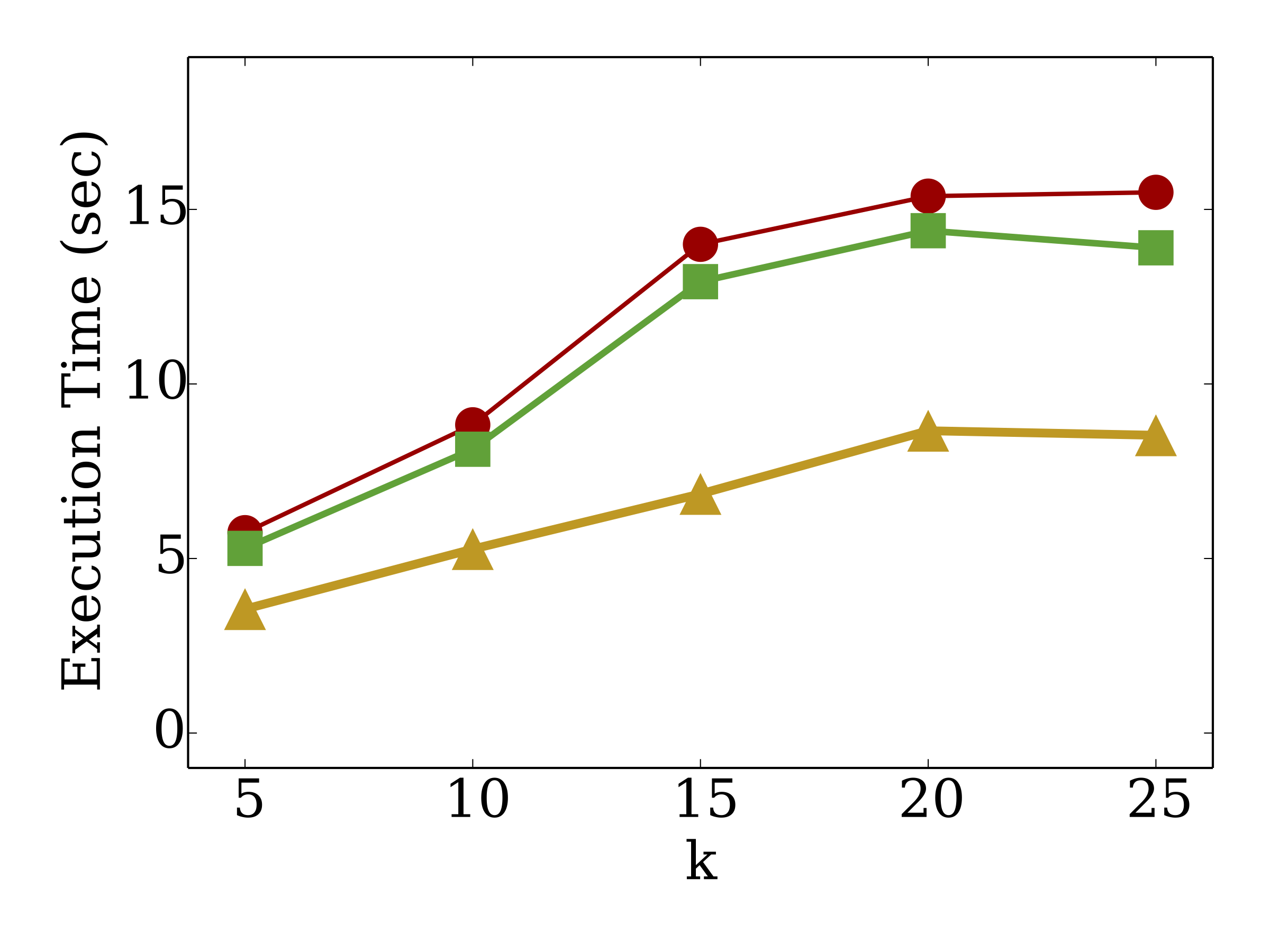}\label{subfig:var_k_crime}}
\subfloat[Crime ($Q_{rk}$)]{\includegraphics[trim=0.5cm 0.5cm 0.5cm 0.5cm, clip, width=0.225\textwidth]{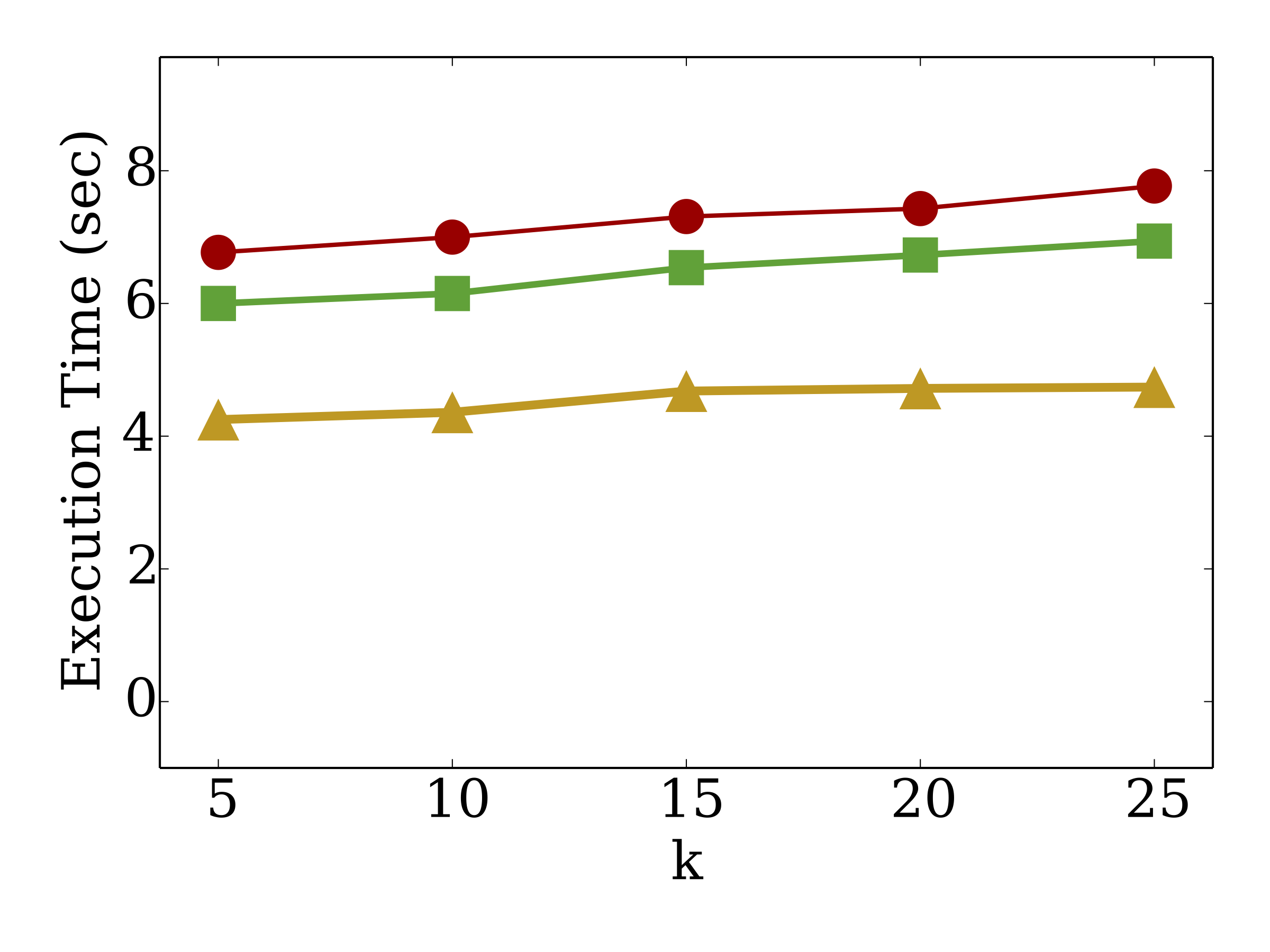}\label{subfig:var_ks_crime}}
\subfloat[$Q_{rr}$]{\includegraphics[trim=0.5cm 0.5cm 0.5cm 0.5cm, clip, width=0.225\textwidth]{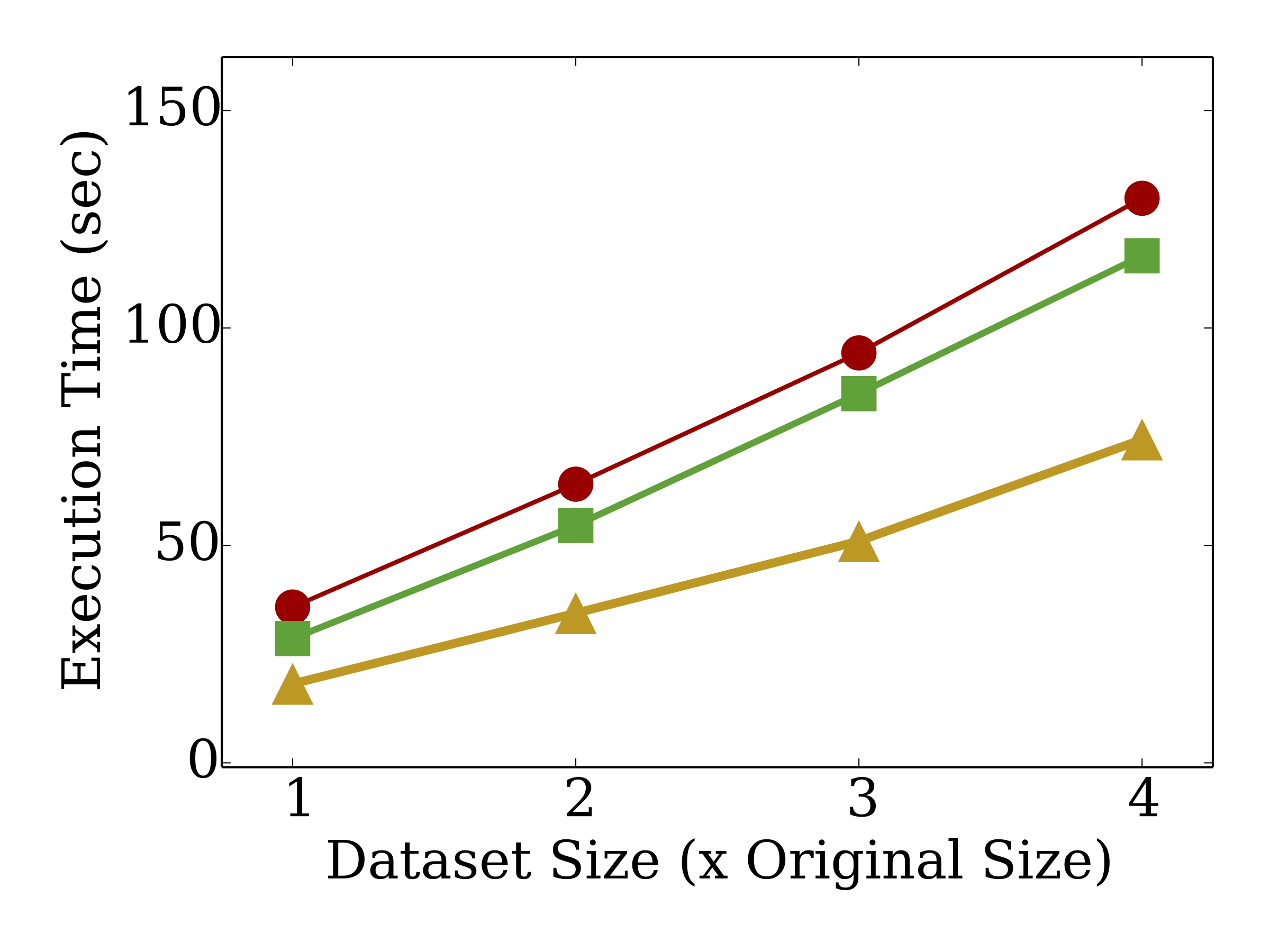}\label{subfig:scalability1}} \\
\vspace{-5pt}
\subfloat[Flickr ($Q_{rr}$)]{\includegraphics[trim=0.5cm 0.5cm 0.5cm 0.5cm, clip, width=0.225\textwidth]{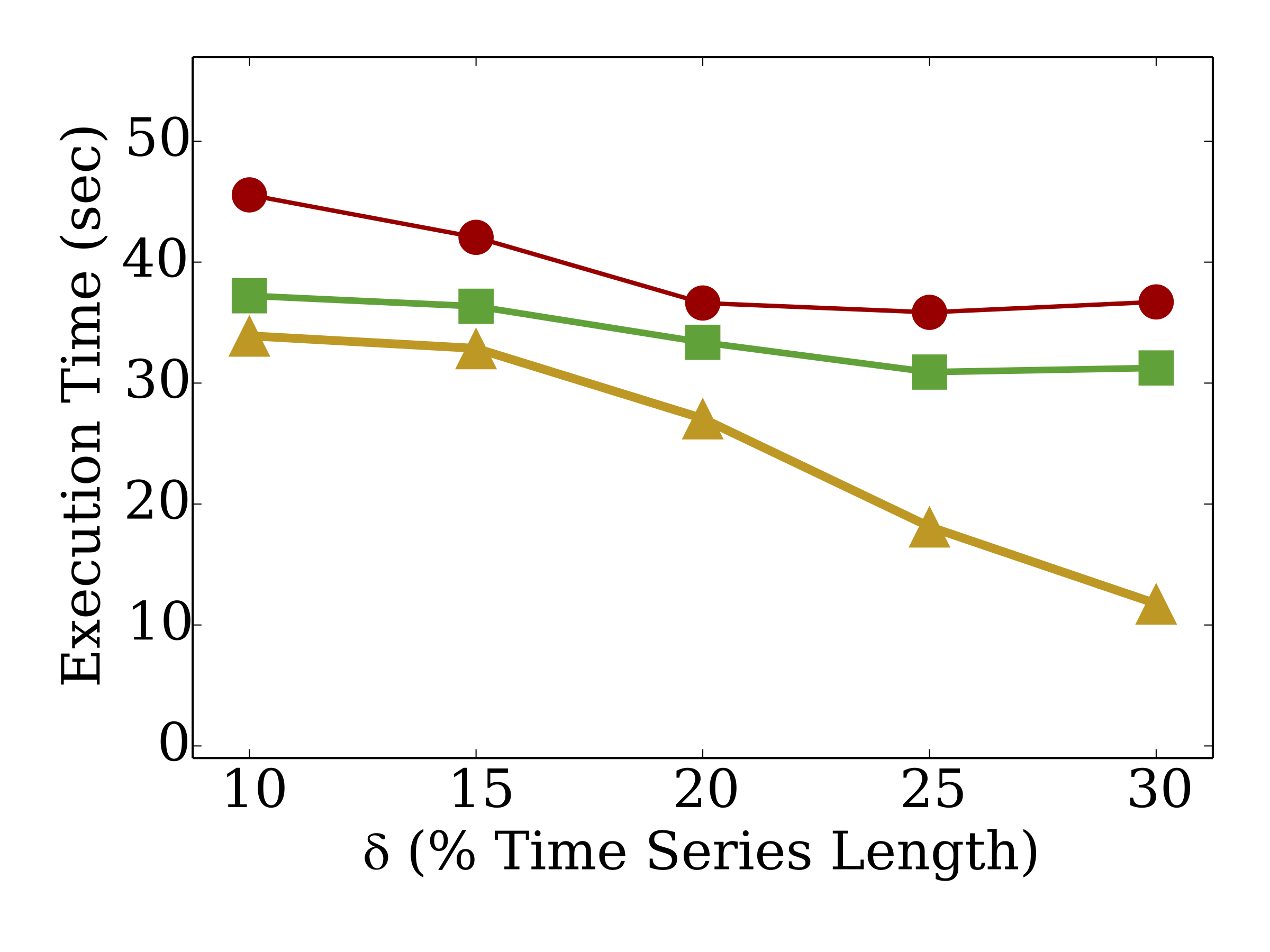}\label{subfig:var_delta_flickr}}
\subfloat[Flickr ($Q_{kr}$)]{\includegraphics[trim=0.5cm 0.5cm 0.5cm 0.5cm, clip, width=0.225\textwidth]{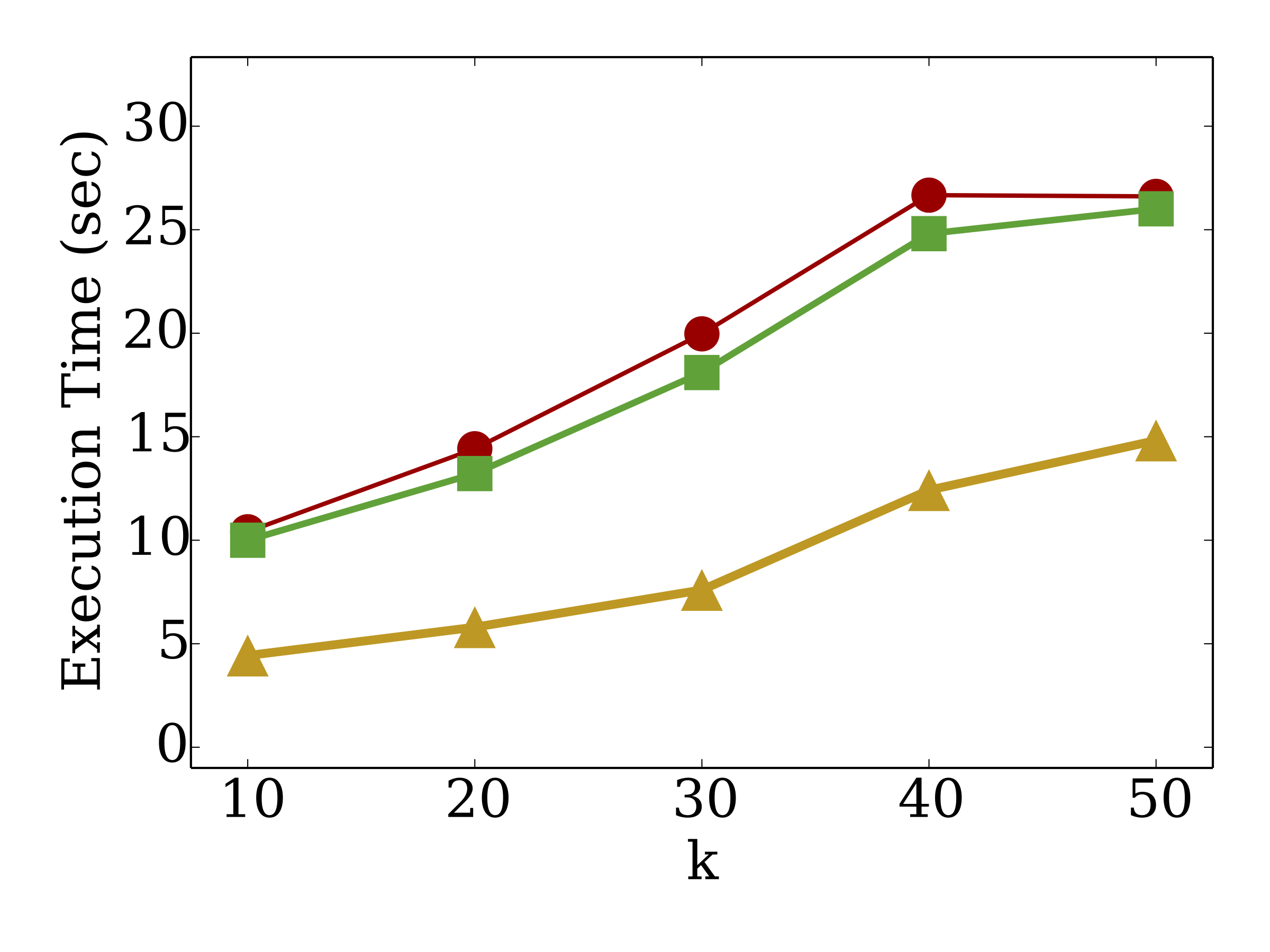}\label{subfig:var_k_flickr}}
\subfloat[Flickr ($Q_{rk}$)]{\includegraphics[trim=0.5cm 0.5cm 0.5cm 0.5cm, clip, width=0.225\textwidth]{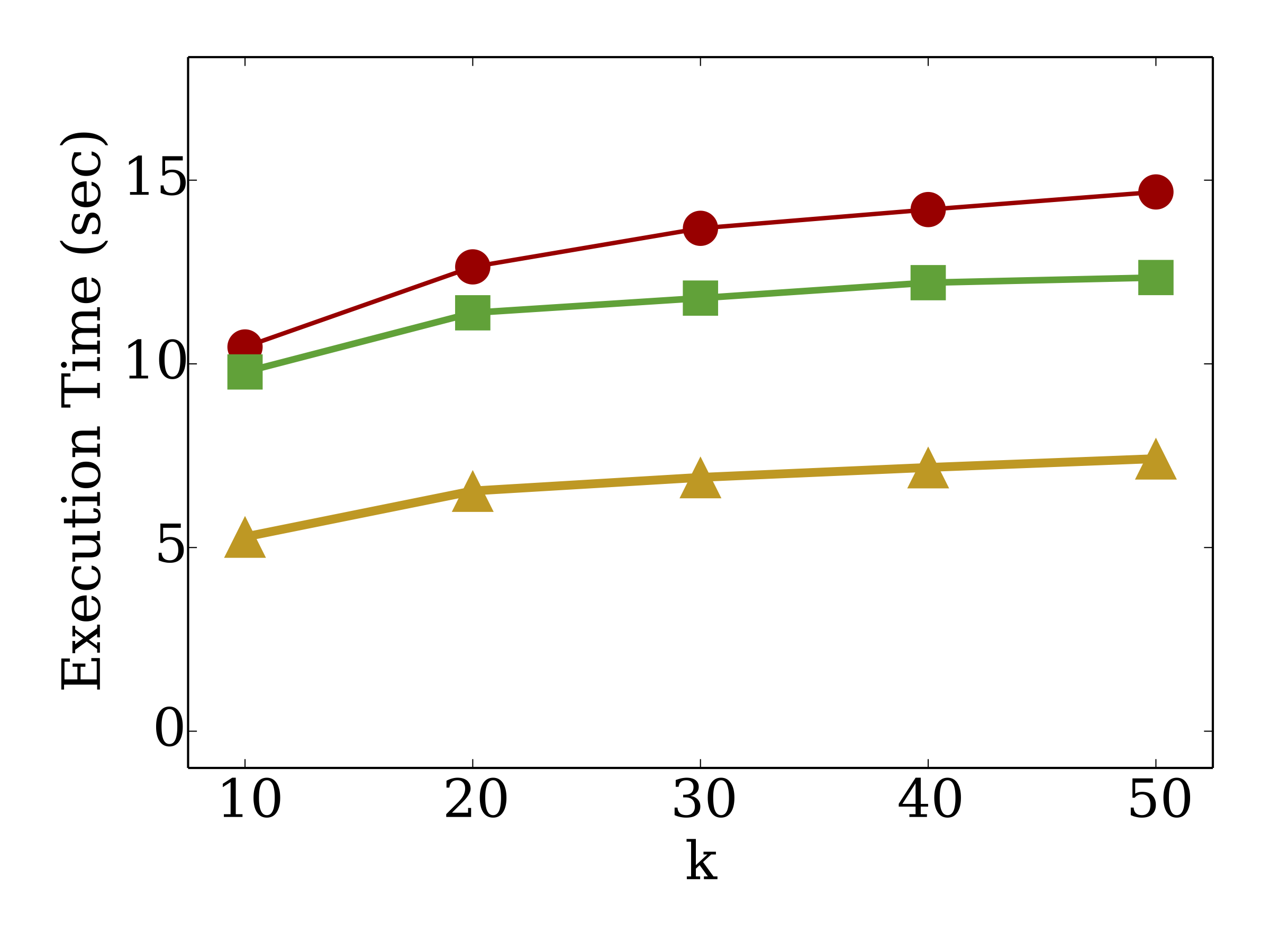}\label{subfig:var_ks_flickr}}
\subfloat[$Q_{kr}$]{\includegraphics[trim=0.5cm 0.5cm 0.5cm 0.5cm, clip, width=0.225\textwidth]{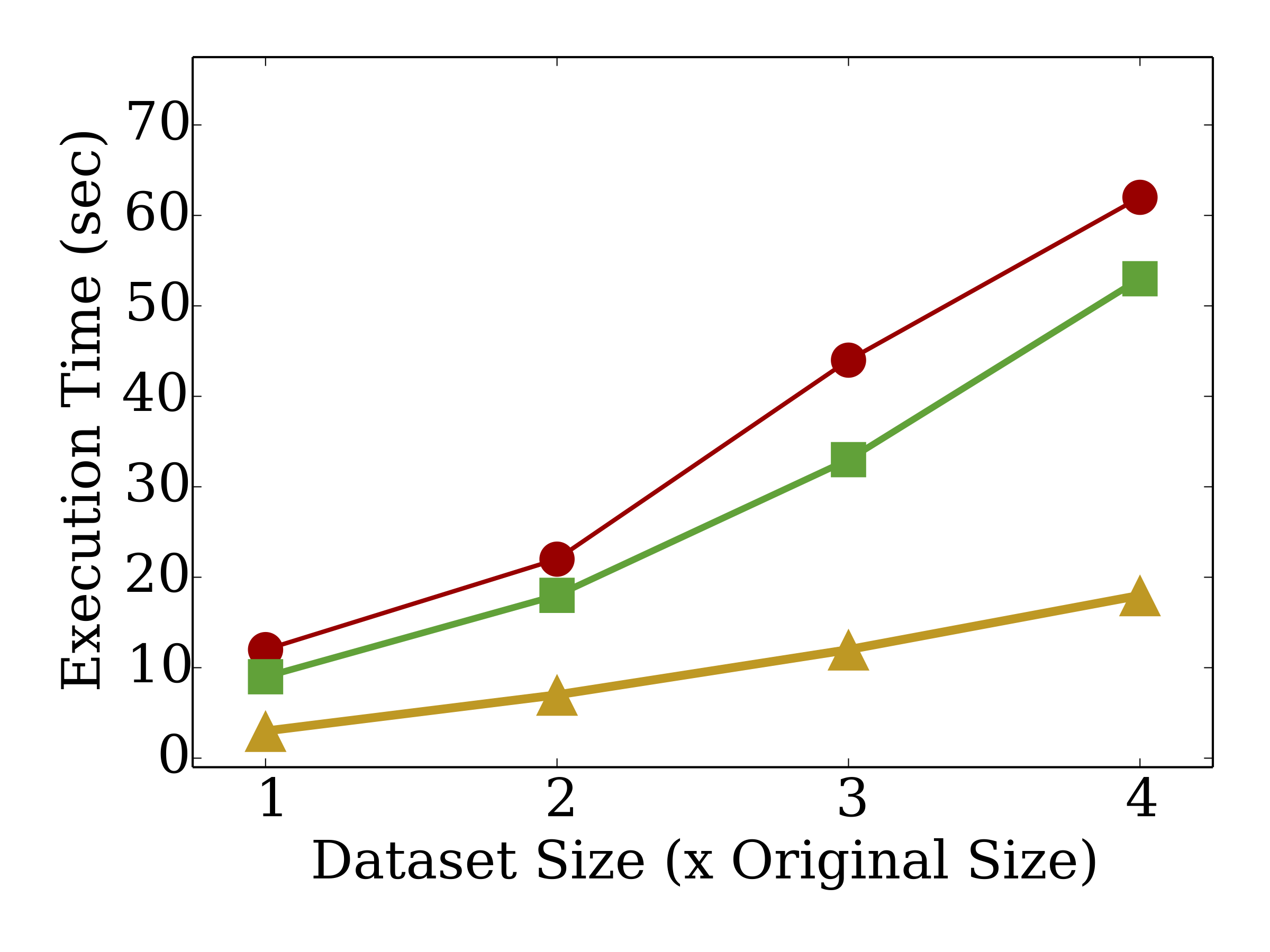}\label{subfig:scalability2}} \\
\vspace{-5pt}
\subfloat[Taxi ($Q_{rr}$)]{\includegraphics[trim=0.5cm 0.5cm 0.5cm 0.5cm, clip, width=0.225\textwidth]{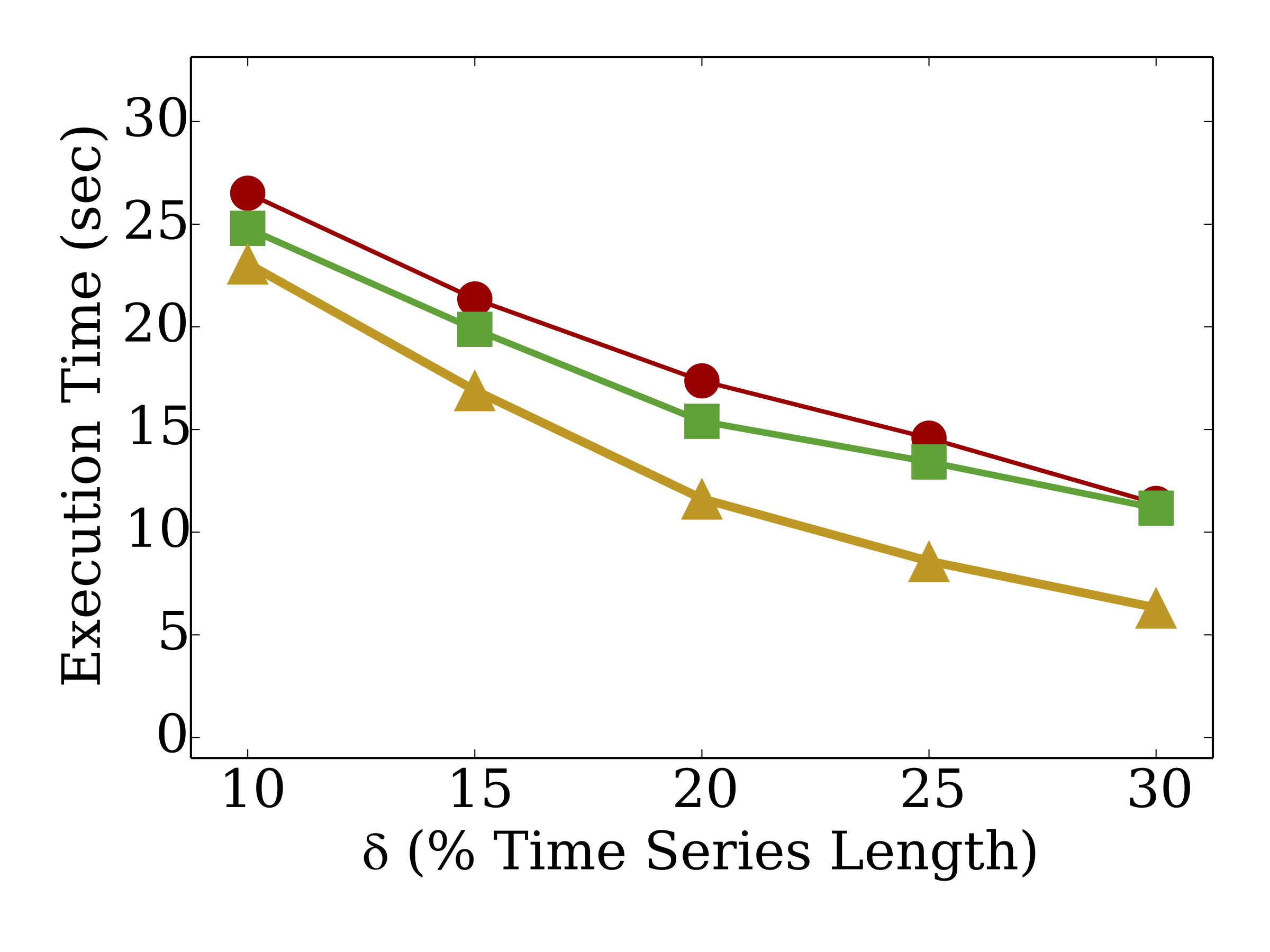}\label{subfig:var_delta_taxi}}
\subfloat[Taxi ($Q_{kr}$)]{\includegraphics[trim=0.5cm 0.5cm 0.5cm 0.5cm, clip, width=0.225\textwidth]{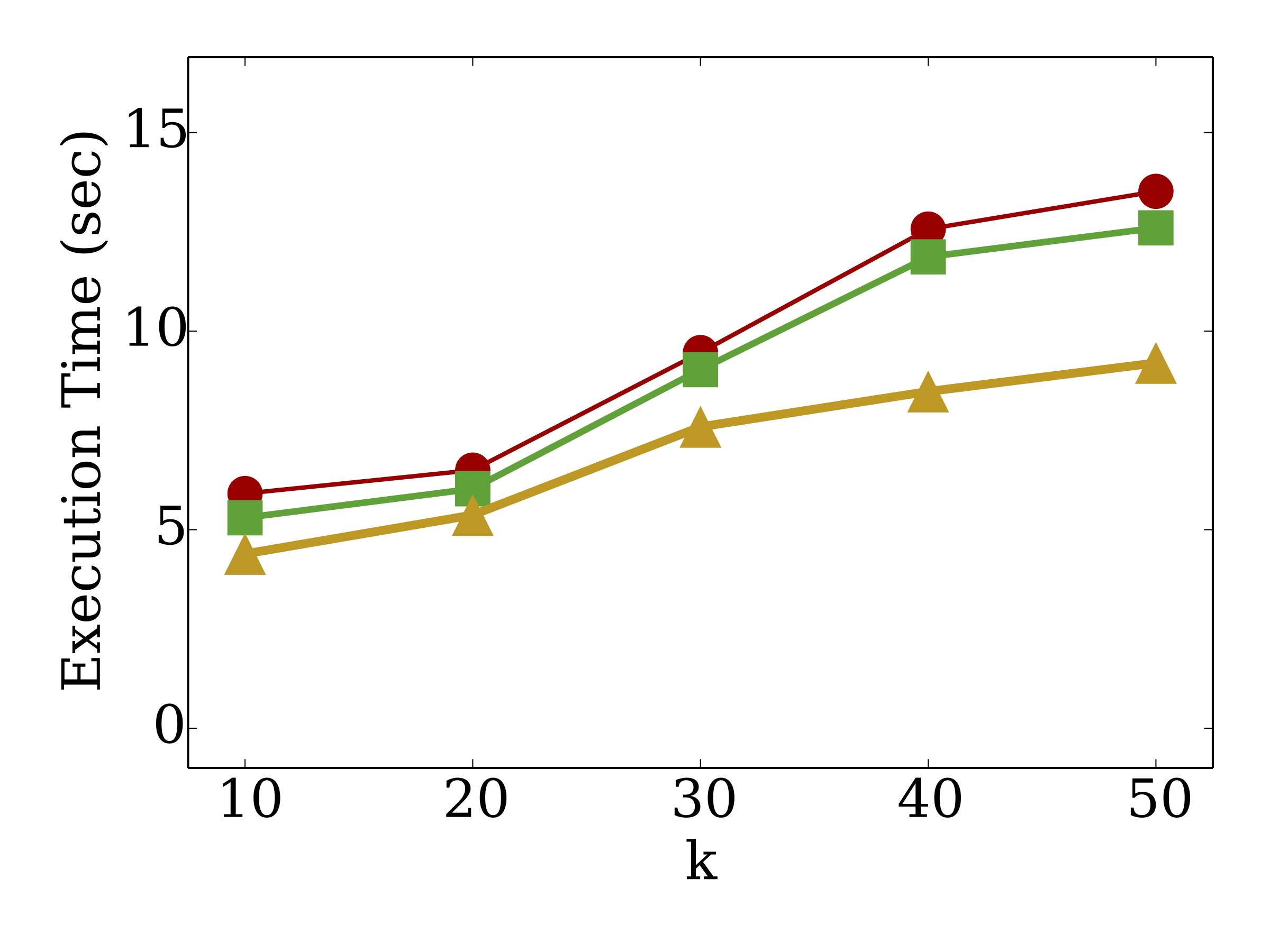}\label{subfig:var_k_taxi}}
\subfloat[Taxi ($Q_{rk}$)]{\includegraphics[trim=0.5cm 0.5cm 0.5cm 0.5cm, clip, width=0.225\textwidth]{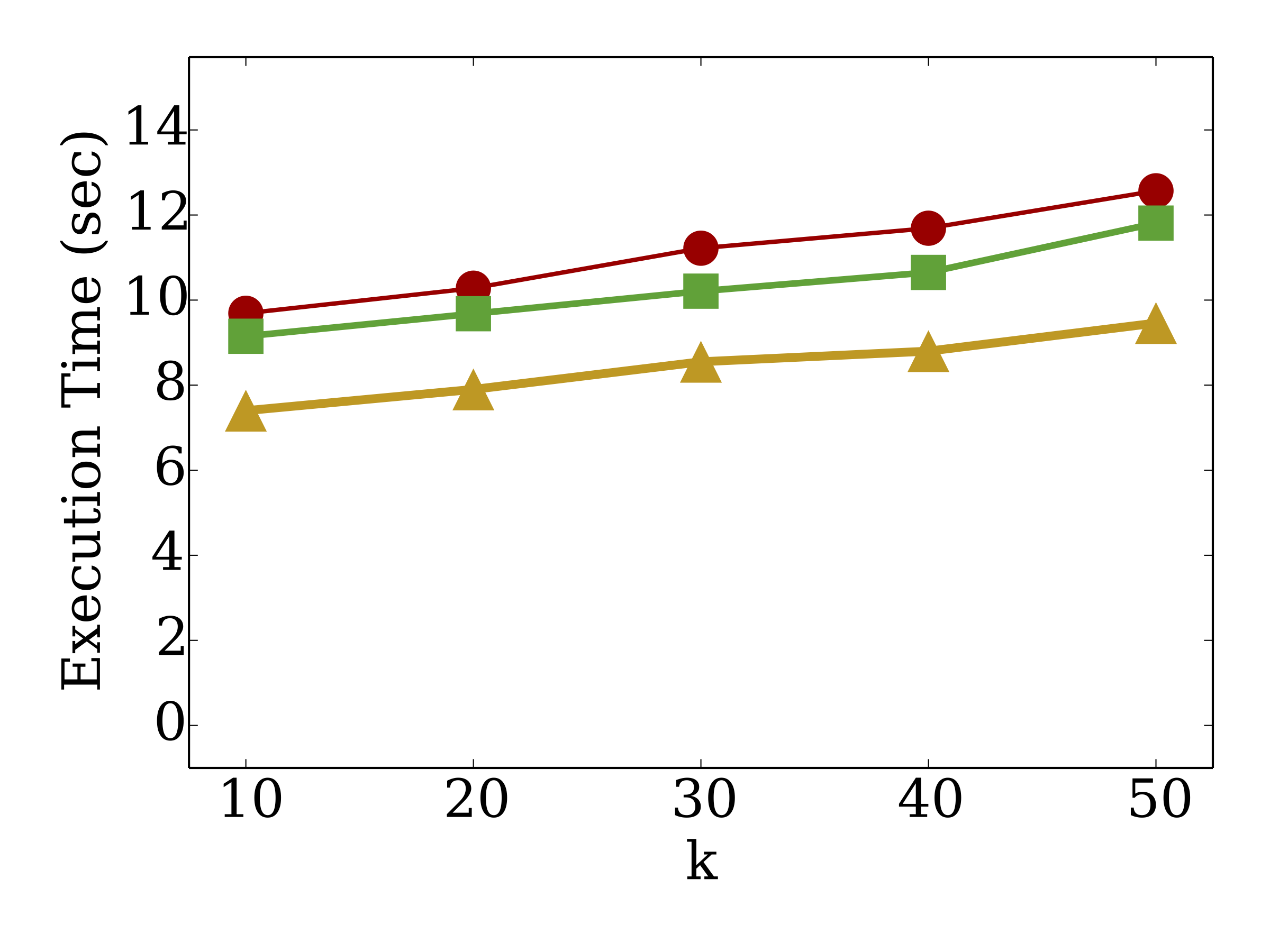}\label{subfig:var_ks_taxi}}
\subfloat[$Q_{rk}$]{\includegraphics[trim=0.5cm 0.5cm 0.5cm 0.5cm, clip, width=0.225\textwidth]{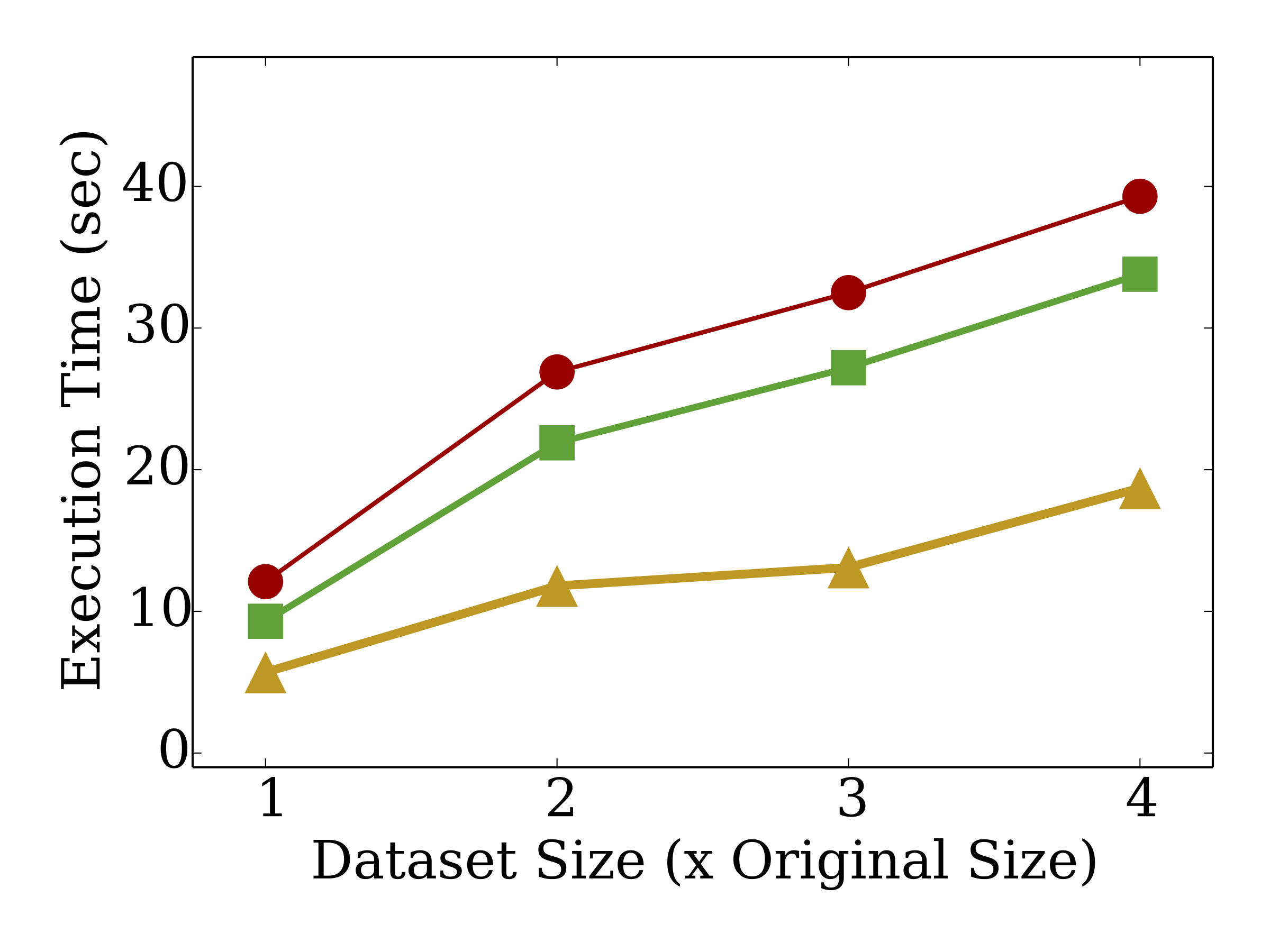}\label{subfig:scalability3}}
\caption{Per column: $Q_{rr}(T_q, \rho, \epsilon, \delta)$ for varying $\delta$ -- $Q_{kr}(T_q, k, \epsilon, \delta)$ for varying $k$ -- $Q_{rk}(T_q, k, \rho)$ for varying $k$ -- Scalability. }
\label{fig:exp}
\end{figure*}

\subsubsection{$Q_{rr}(T_q, \rho, \epsilon, \delta)$}
Figure \ref{fig:query1a} illustrates the query performance for varying thresholds $\rho$ and $\epsilon$ and the first column of Figure \ref{fig:exp} for varying $\delta$, on all three datasets. It is apparent that the \sbtsr with the checkpoint approach outperforms the rest in all cases. Its superior pruning power is attributed to the segmentation, which yields tighter bounds within the nodes and consequently less disk accesses. The sweep line and checkpoint methods over \btsr perform similarly in all cases. Both methods access the same nodes, but the checkpoint approach needs to examine significantly less values across time in order to determine local similarities. However, since all local similarity calculations take place in-memory, computation cost does not make a big difference, compared to the lesser node accesses required with the \sbtsr. 

More specifically, for the {\em crime} dataset, relaxing $\rho$ (Figure \ref{subfig:var_epsSP_crime}) has a negative impact on all three methods as more nodes have to be accessed and pruning depends mostly on the $\epsilon$ value. \sbtsr increasingly outperforms the rest as $\rho$ increases, due to its more aggressive pruning on local similarity. For the case of increasing $\epsilon$ (Figure \ref{subfig:var_epsTS_crime}), the result is the opposite, as this way the parameter is relaxed and more nodes get accessed. For very large $\epsilon$ values, pruning is solely based on spatial distance and all approaches perform similarly. Finally, increasing $\delta$ (Figure \ref{subfig:var_delta_crime}) also increases the difference in performance among the three approaches, while it also reduces the average query response time. This is due to large numbers of subsequences qualifying for small $\delta$ values, resulting in more node accesses. As $\delta$ increases, pruning is more rapidly improved in the case of \sbtsr due to its tighter bounds.

The results are similar but with larger differences for the {\em Flickr} dataset (Figures \ref{subfig:var_epsSP_flickr}, \ref{subfig:var_epsTS_flickr} and \ref{subfig:var_delta_flickr}). Intuitively, the less periodicity in a dataset, the more the benefit from segmentation; if the time series in the dataset exhibit periodicity, the bounds that will occur from applying $k$-means clustering on the whole sequences will be relatively tighter than otherwise. The Flickr dataset, due to its nature, is more random than the crime dataset, which justifies the larger differences. This explanation is also supported by the results for the {\em taxi} dataset, illustrated in Figures \ref{subfig:var_epsSP_crime}, \ref{subfig:var_epsTS_crime} and \ref{subfig:var_delta_crime}. Despite a similar behavior in varying all thresholds, the differences in average query response time among the different approaches are smaller than in the crime and Flickr datasets, due to the high daily periodicity of taxi drop-offs.


\subsubsection{$Q_{kr}(T_q, k, \epsilon, \delta)$}
Figures \ref{subfig:var_k_crime}, \ref{subfig:var_k_flickr} and \ref{subfig:var_k_taxi} depict the results for the $Q_{kr}(T_q, k, \epsilon, \delta)$ query for the three datasets. As $k$ increases, more nodes have to be traversed in order to fetch the additional results, and the execution time increases for all methods. Nevertheless, \sbtsr still clearly outperforms the other two algorithms.


\subsubsection{$Q_{rk}(T_q, k, \rho)$}
Finally, Figures \ref{subfig:var_ks_crime}, \ref{subfig:var_ks_flickr} and \ref{subfig:var_ks_taxi} depict the results for the $Q_{rk}(T_q, k, \rho)$ query. In this case, the performance deterioration as $k$ increases is less abrupt, especially for the crime dataset, as usually the top-$k$ results are spatially closely located and are retrieved quickly. Again, the largest and smallest differences are spotted on the Flickr and taxi datasets, respectively.

\subsection{Scalability}
\label{subsec:scalability}
We performed a scalability evaluation for all three queries using the Flickr-based synthetic datasets, again measuring the average query response time for the same query workload. The results for increasing dataset size (up to four times) are depicted in Figure \ref{fig:exp}. In all cases, the \sbtsr-based approach scales better, especially in the top-$k$ queries (Figures \ref{subfig:scalability2} and \ref{subfig:scalability3}), where the larger difference observed in Figures \ref{subfig:var_k_flickr} and \ref{subfig:var_ks_flickr} is further augmented.

\section{Conclusions}
\label{sec:conclusions}

We have studied three variants of hybrid queries on geolocated time series, involving both range and top-$k$ search, and combining spatial distance with local time series similarity. The latter allows to measure similarity of time series over subsequences instead of their entire length, and thus enables the identification of more fine-grained trends and patterns. The queries are evaluated by hybrid index structures, in order to allow for simultaneous pruning by both criteria. We first discuss query evaluation using the previously proposed \btsr, and then we further extend it to derive the \sbtsr which exhibits even better performance, by using temporal segmentation of time series to derive tighter bounds. Our evaluation against several real-world datasets has shown that \sbtsr can compute results much faster for all query variants.



\bibliographystyle{abbrv}
\bibliography{References}

\end{document}